\documentclass[fleqn,usenatbib]{mnras}

\usepackage{newtxtext,newtxmath}

\usepackage[T1]{fontenc}

\DeclareRobustCommand{\VAN}[3]{#2}
\let\VANthebibliography\thebibliography
\def\thebibliography{\DeclareRobustCommand{\VAN}[3]{##3}\VANthebibliography}

\usepackage{graphicx}	
\usepackage{amsmath}	

\title[Chemical Composition of Protoplanetary Disks]{Inside-Out Planet Formation. VII. Astrochemical Models of Protoplanetary Disks and Implications for Planetary Compositions}

\author[Cevallos Soto et al.]{
Arturo Cevallos Soto,$^{1}$\thanks{E-mail: cev.arturo@gmail.com}
Jonathan C. Tan,$^{1,2}$
Xiao Hu$^{2}$
Chia-Jung Hsu$^{1}$
Catherine Walsh$^{3}$
\\
$^{1}$Department of Space, Earth and Environment, Chalmers University of Technology, Gothenburg SE-412 96, Sweden\\
$^{2}$Department of Astronomy, University of Virginia, Street Address, Charlottesville, VA 22904, USA\\
$^{3}$School of Physics and Astronomy, University of Leeds, Leeds LS2 9JT, UK}

\date{Accepted XXX. Received YYY; in original form ZZZ}

\pubyear{2022}

\begin{document}
\label{firstpage}
\pagerange{\pageref{firstpage}--\pageref{lastpage}}
\maketitle

\begin{abstract}
Inside-Out Planet Formation (IOPF) proposes that the abundant systems of close-in Super-Earths and Mini-Neptunes form {\it in situ} at the pressure maximum associated with the Dead Zone Inner Boundary (DZIB). We present a model of physical and chemical evolution of protoplanetary disk midplanes that follows gas advection, radial drift of pebbles and gas-grain chemistry to predict abundances from $\sim300\:$au down to the DZIB near 0.2~au. We consider typical disk properties relevant for IOPF, i.e., accretion rates $10^{-9}< \dot{m}/ (M_\odot\:{\rm{yr}}^{-1} )<10^{-8}$ and viscosity parameter $\alpha=10^{-4}$, and evolve for fiducial duration of $10^5\:$yrs. For outer, cool disk regions, we find that C and up to $90\%$ of O nuclei start locked in CO and $\rm O_2$ ice, which keeps abundances of $\rm CO_2$ and $\rm H_2O$ one order of magnitude lower. Radial drift of icy pebbles is influential, with gas-phase abundances of volatiles enhanced up to two orders of magnitude at ice-lines, while the outer disk becomes depleted of dust. Disks with decreasing accretion rates gradually cool, which draws in icelines closer to the star. At $\lesssim1\:$au, advective models yield water-rich gas with C/O ratios $\lesssim0.1$, which may be inherited by atmospheres of planets forming here via IOPF. For planetary interiors built by pebble accretion, IOPF predicts volatile-poor compositions. However, advectively-enhanced volatile mass fractions of $\sim10\%$ can occur at the water ice line.
\end{abstract}




\section{Introduction} \label{sec:intro}

The discovery by NASA's {\it Kepler} mission of an abundant class of Systems with Tightly-packed Inner Planets (STIPs) has challenged traditional planet formation models, which typically involved planets forming in the outer disk and then migrating to inner disk locations. As reviewed by \cite{tan2015overview}, Inside-Out Planet Formation (IOPF) \citep[][]{CT14} 
proposes that planets form sequentially \textit{in situ} from the inside-out via creation of successive pebble-rich rings fed from a continuous stream of pebbles, drifting inwards due to gas drag. At the onset of IOPF the disk has an inner magnetorotational instability (MRI)-active zone extending to between $\sim0.1$ to 1~AU, depending on the accretion rate at this time. Beyond this extends the MRI-inactive zone, i.e., ``dead zone''. A pressure maximum exists in the transition between these two zones, i.e., at the Dead Zone Inner Boundary (DZIB). Pebbles drifting towards the star will be trapped at the pressure maximum, eventually forming a ring, which would then collapse into a protoplanet, via a variety of processes \citep[see, e.g.,][]{Hu_2015,cai2021insideout}. When the protoplanet becomes massive enough it will open a gap in the disk, pushing the pressure maximum outwards by at least several Hill radii and starting the process anew. 

While this process is taking place, the protoplanetary disk is expected to be undergoing physical and chemical evolution \citep[see, e.g.,][]{Walsh_2010,Walsh_2015,Booth_2017,Booth_2019}.
Following this evolution accurately is crucial for understanding the bulk composition of planetary cores, which may be mostly set by pebble accretion in the IOPF model, and of the composition of their atmospheres, which may be dominated by residual gas accretion from the disk.

Gaseous components undergo gas-phase reactions, but also physical (e.g., freeze-out) and chemical interactions with the surfaces of dust grains. Further chemical reactions are confined to species in the dust grain ice mantles, especially the build up of more complex organic molecules (COMs). Dust grains are themselves expected to evolve by growing in size, i.e., by sweeping-up smaller grains, coagulation with similar sized grains and growth of ice mantles. Larger grains begin to decouple from the gas leading to vertical settling and radial inward drift due to headwind gas drag.
While drifting inwards, pebbles carry volatile species frozen out onto their surfaces to higher temperature regions. These ice species sublimate after crossing their respective ice-lines leading to both physical and chemical changes \citep[see, e.g.,][]{Booth_2017,Booth_2019,Molyarova_2021}. This makes them available for further high temperature gas-phase chemical processes. Gas advection provides a slower, but steady, supply of new gas species to inner regions close to the DZIB. With disk lifetimes potentially on the order of Myrs, there is likely to be enough time for drastic changes in the chemical composition of the disk due to advective transport of gas and pebbles.

Examples of previous works on modeling the chemistry of protoplanetary disks include those of \cite{Eistrup_2016} and \cite{Eistrup_2018}, who explored detailed chemical models of disk midplanes using a gas-grain network with 668 species and 8764 reactions. They found cosmic-ray induced chemistry and gas-grain reactions were important for the resulting dominance of $\rm H_2O$ and $\rm CO_2$ ices in the outer disk, at the expense of CO ice and $\rm CH_4$ gas. Initial choices for abundances were found to have an impact on the chemical makeup of the disk even after 1 Myr, including the overall C/O ratios in gas and ice species. However, in these studies the physical processes of gas and dust transport and grain growth were neglected.

Additionally, \cite{Booth_2017} and \cite{Booth_2019} have investigated coupling chemical evolution and matter transport in protoplanetary disks, demonstrating the influence of gas advection and pebble drift on the resulting composition of planets. They found that outer disk pebbles will influence the composition of inner regions by transporting ices of volatile species inwards as they migrate. These different ices will sublimate at their respective ice-lines and enrich the local gas. However, this investigation was done using a relatively simple chemical network with 136 species and 1485 reactions.

Here we present a model coupling the chemical evolution in a disk midplane with the disk evolution due to gas advection and pebble drift, as well as including cases with a secular decline in the global accretion rate. We make use of an extensive gas-grain network first developed by \cite{Walsh_2010} and also used by \citet[][]{Eistrup_2016,Eistrup_2018}. We will pay special attention to those regions in the disk close to the DZIB (between 0.1 and 1 AU), where the pressure gradients and radial drift speeds are high. We focus on single-size grain/pebble models at any given local location in a disk and exclude turbulent diffusion \citep[e.g.,][]{Schoonenberg_2017} from the dynamics, assuming its impact on the evolution to be small, which is expected to be reasonable in the dead zone regions of the disk.

The remainder of this work is structured as follows. In \S\ref{sec:method} we describe the physical characteristics of the model, the chemical model and the coupling of the chemical evolution and transport mechanisms. In \S\ref{sec:res} we present the results concerning the evolution of the main volatile abundances in progressively more sophisticated and realistic disk models, while also reviewing each model's C/O ratio and dust-to-gas mass ratio. In \S\ref{sec:dis} we discuss some key points concerning the implications of this work for IOPF, as well as the key limitations of the modeling. In \S\ref{sec:con} we present our conclusions. Supplementary results are presented in Appendix \S\ref{sec:appendix} and \S\ref{sec:appendix_1Myr}.

\section{Methods}\label{sec:method}

\subsection{Disk physical properties}\label{subsec:disk}

We model the disk structure with an $\alpha$-disk framework \citep{1973AA....24..337S}, following previous IOPF works by \cite{CT14} and \cite{Hu_2018}. The model, which aims to specify disk midplane conditions in particular, assumes the disk is a vertically thin and in a steady state described by a constant accretion rate at all radial locations. The thermal structure of the inner regions is dominated by active accretion heating, while that of the outer regions is mostly set by passive heating from the star. Thus the main properties of the disk are controlled by the parameters of accretion rate ($\dot{m}$), stellar mass ($m_{*}$), stellar luminosity ($L_*$), and the dimensionless viscosity parameter ($\alpha$). From \cite{Hu_2018}, the following equations are use to solve for disk structure (with the cases for the temperature profiles for the active and passive regions of the disk described below):
\begin{equation}
\label{eqn:Tr}
    T = T(r)
\end{equation}
\begin{equation}
    c_s(r) = (\gamma k_B T / \mu)^{1/2}
\end{equation}
\begin{equation}
    h(r)/r = c_s / v_K
\end{equation}
\begin{equation}
    \nu(r) = \alpha c_s h
\end{equation}
\begin{equation}
    \Sigma_g(r) = \dot{m}/(3 \pi \nu)
\end{equation}
\begin{equation}
\label{eqn:rho}
    \rho(r) = \Sigma_g / (h \sqrt{2 \pi}),
\end{equation}
where $c_s$ is the midplane sound speed, $\gamma=1.4$ is the power law exponent of the barotropic equation of state $P = K \rho^{\gamma}$, $k_B$ is Boltzmann's constant, $\mu = 2.33 m_{\text{H}} = 3.90 \times 10^{-24}$ g is the mean particle mass (i.e., assuming $n_{\rm He}=0.2 n_{\rm H_2}$), $h$ is the disk vertical scale height, $v_K$ is the local Keplerian speed, $\nu$ is the viscosity, $\Sigma_g$ is the gas mass surface density, $\rho$ is the midplane gas density, and $n$ is the midplane number density of gas particles. The disk orbits around a young, Sun-like star with a stellar mass of $m_{*} = 1\:M_{\odot}$, a stellar radius $r_{*} = 3\:R_{\odot}$, an effective temperature of $T_*=4500\:$K and thus a luminosity of $L_*=3.3\:L_\odot$.

From \cite{CT14}, for the ``active'' part of the disk, i.e., dominated by local accretion heating, the radial temperature profile, $T(r)$, is described by:
\begin{eqnarray}
\label{eq:temp}
    T(r) = \frac{3^{1/5}}{2^{7/5} \pi^{2/5}}\left( \frac{\mu}{\gamma k_B} \right)^{1/5} \left( \frac{\kappa}{\sigma_{\text{SB}}} \right)^{1/5} \nonumber \\
    \times \alpha^{-1/5}(G m_{*})^{3/10}(f_r \dot{m})^{2/5} r^{-9/10} \text{ K} \nonumber \\
    \rightarrow 290 \gamma_{1.4}^{-1/5} \kappa_{10}^{1/5} \alpha_{-4}^{-1/5} m_{*,1}^{3/10}(f_r \dot{m}_{-9})^{2/5} r_{\rm au}^{-9/10}\:{\rm K},
\end{eqnarray}
where $\kappa$ is the opacity (we use tabulated opacities from \citeauthor{Zhu_2009} \citeyearpar{Zhu_2009}, while the fiducial numerical evaluation indicated by $\rightarrow$ adopts $\kappa=10\:{\rm cm^2\:g^{-1}}$ along with other fiducial parameter normalisations), $\sigma_{\text{SB}}$ is the Stefan-Boltzmann constant, $G$ is the gravitational constant, $f_r \equiv 1 - \sqrt{r_{*}/r}$ and $r_{\rm au} \equiv r/(1 {\rm au})$. In the outer regions of the disk, beyond the transition region where the irradiation temperature equals the accretion heating, the disk is considered to be ``passive'' and has its temperature switched to being described by \citep[see][]{Hu_2018}:
\begin{equation}
    T(r) = 172 r_{\rm AU}^{-3/7} \:{\rm K}.
\end{equation}

In the active, inner region, the gas mass surface density profile is given by:
\begin{eqnarray}
\label{eq:surf_dens}
    \Sigma_g & = & \frac{2^{7/5}}{3^{6/5} \pi^{3/5}}\left( \frac{\mu}{\gamma k_B} \right)^{4/5} \left( \frac{\kappa}{\sigma_{\text{SB}}} \right)^{-1/5} \nonumber \\
    & \times & \alpha^{-4/5} (G m_{*})^{1/5}(f_r \dot{m})^{3/5} r^{-3/5} \nonumber \\
    & \rightarrow & 880 \gamma_{1.4}^{-4/5} \kappa_{10}^{-1/5} \alpha_{-4}^{-4/5} m_{*,1}^{1/5} \nonumber \\ 
    & & (f_r \dot{m}_{-9})^{3/5} r_{\rm au}^{-3/5}\:{\rm g\:cm}^{-2},
\end{eqnarray}
while in the outer, passive regime it is given by:
\begin{eqnarray}
\label{eq:surf_dens_passive}
    \Sigma_g & = & \frac{1}{516 \pi} (1 \:{\rm au})^{-3/7} \frac{\mu}{\gamma k_B} \nonumber \\
    & \times &  \alpha^{-1} (G m_{*})^{1/2} \dot{m} r^{-15/14} \nonumber \\
    & \rightarrow &  11.2 \gamma_{1.4}^{-1} \alpha_{-4}^{-1} m_{*,1}^{1/2} \dot{m}_{-9} r_{\rm 100 au}^{-15/14} \:{\rm g\:cm}^{-2}.
\end{eqnarray}

Along with temperature, the number density of gas particles, $n$, is crucial for controlling astrochemical reaction rates. For convenience we express this in terms of number density of H nuclei, $n_{\rm H}$, which in the approximate limiting case of H in fully molecular form, $n_{\rm He}=0.1 n_{\rm H}$, and negligible other species is related to $n$ via $n_{\rm H} = (5/3) n$. 
Thus, for the active region:
\begin{eqnarray}
\label{eq:num_dens_active}
    n_{\rm H} & = & \frac{5}{3}\frac{2^{8/5}}{3^{13/10} \pi^{9/10}} \frac{\mu^{1/5}}{\left( \gamma k_B \right)^{6/5}} \left( \frac{\kappa}{\sigma_{\text{SB}}} \right)^{-3/10} \nonumber \\
    & \times & \alpha^{-7/10} (G m_{*})^{11/20}(f_r \dot{m})^{2/5} r^{-33/20} \nonumber \\
    & \rightarrow & 2.43 \times 10^{14} \gamma_{1.4}^{-6/5} \kappa_{10}^{-3/10} \alpha_{-4}^{-7/10} m_{*,1}^{11/20} \nonumber \\ 
    & & \times (f_r \dot{m}_{-9})^{2/5} r_{\rm au}^{-33/20} \:{\rm cm}^{-3},
\end{eqnarray}
while for the passive region:
\begin{eqnarray}
\label{eq:num_dens_passive}
    n_{\rm H} & = & \frac{5}{3}\frac{1}{1032 \sqrt{86} \pi^{3/2}} (1 \:{\rm au})^{-9/14} \frac{\mu^{1/2}}{\left( \gamma k_B \right)^{3/2}} \nonumber \\
    & \times &  \alpha^{-1} G m_{*} \dot{m} r^{-6/7} \nonumber \\
    & \rightarrow &  1.11 \times 10^{10} \gamma_{1.4}^{-1} \alpha_{-4}^{-1} m_{*,1}^{1/2} \dot{m}_{-9} r_{\rm 100 au}^{-15/14} \:{\rm cm}^{-3}.
\end{eqnarray}

\begin{figure*}
    \centering
    \includegraphics[width=1.0\textwidth]{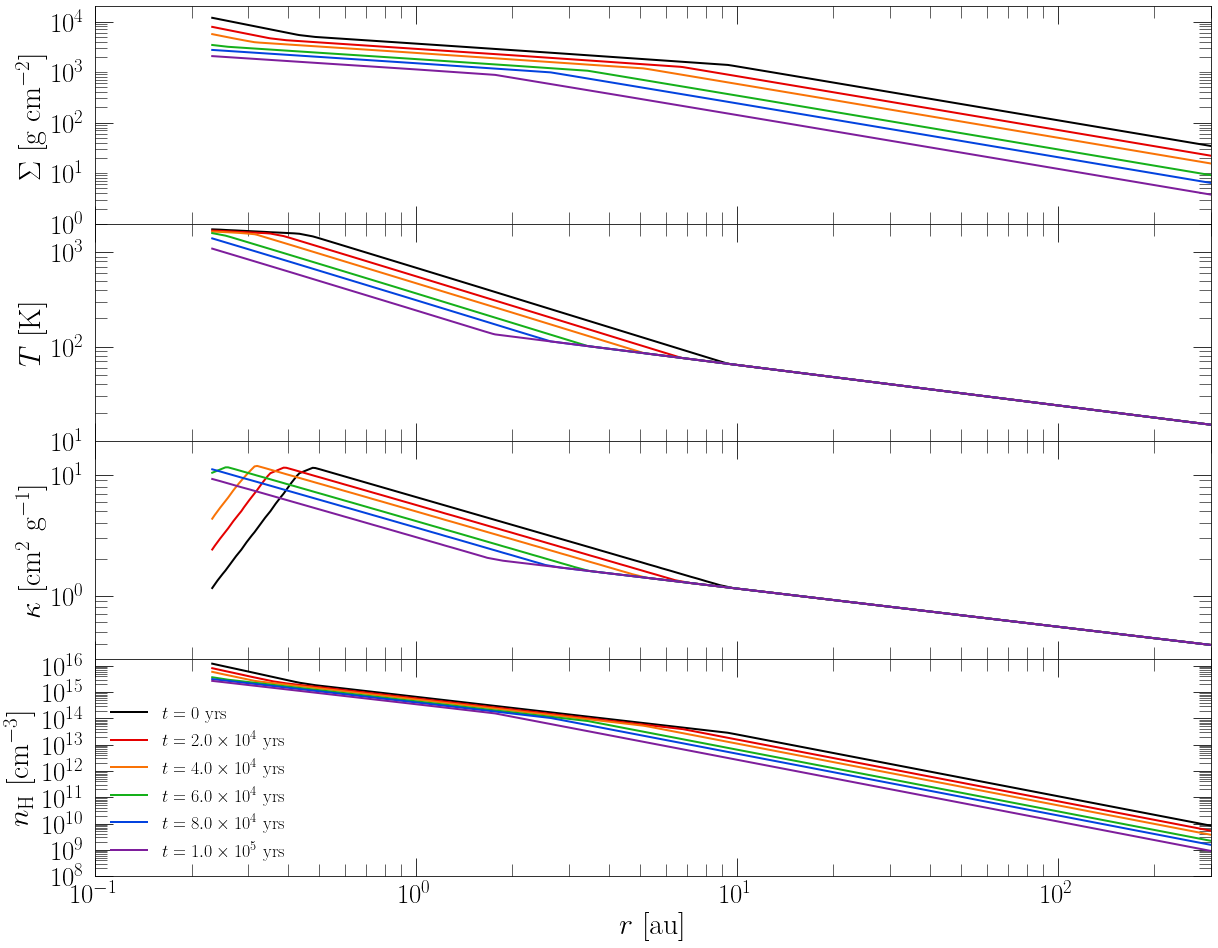}
    \caption{Radial profiles of disk properties, showing an evolutionary sequence for an exponentially decaying accretion rate from $\dot{m} = 10^{-8}\:M_{\odot}\:{\rm yr}^{-1}$ to $\dot{m} = 10^{-9}\:M_{\odot}\: {\rm yr}^{-1}$ in $10^5$ years. From top to bottom, the rows show: gas mass surface density ($\Sigma$), midplane temperature ($T$), midplane opacity ($\kappa$), and midplane number density of H nuclei ($n_{\rm H}$).}
    \label{fig:param}
\end{figure*}{}

Based on observations of T-Tauri stars and transition disks \citep[see, e.g.,][]{Alcal__2013,Manara_2014}, $\dot{m}$ is expected to have values between $10^{-8}$ and $10^{-10}\:M_{\odot}\:{\rm yr}^{-1}$. For the Dead Zone, according to the simulations of \cite{Dzyurkevich_2010}, the typical values of $\alpha$ fall in the range of $\sim 10^{-4}$ to $\sim 10^{-3}$. In the MRI-active region $\alpha$ increases to values up to $\sim 10^{-2}$ to $\sim 0.1$. However, this active region, which occurs initially when $T>1200\:$K, is ignored in our modeling; i.e., we will focus only on the disk from large scales down to the DZIB. Thus, in our models we will adopt a fixed value of $\alpha=10^{-4}$ everywhere in the disk. For accretion rates, we will first consider models that either have fixed values of $\dot{m}=10^{-8}\:M_{\odot}\:{\rm yr}^{-1}$ or $\dot{m}=10^{-9}\:M_{\odot}\:{\rm yr}^{-1}$. In the fiducial case we will consider the evolution of the dust/pebbles and chemistry over a period of $10^5\:$yr, which we will see is long enough for static disk models to reach near chemical equilibrium.
We will also consider a model with a time-evolving accretion rate that is described via exponential decay \citep[e.g.,][]{bitsch2015growth,Hu_2018}:
\begin{equation} \label{eq:mdot}
    \dot{m} = \dot{m}_0 e^{-t/t_0},
\end{equation}
with $\dot{m}_0=10^{-8}\:M_{\odot}\:{\rm yr}^{-1}$ and $t_0=4.34 \times 10^4$ years, so that the accretion rate decreases by a factor of $10$ in $10^5$ years. However, as discussed in Appendix~\ref{sec:appendix_1Myr}, the timescale of this decay is quite uncertain and there we present results of a model with a longer decay time of $10^6$ years.

In our disk models the radial grid consists of 96 points spaced equally in log $r$ from an inner disk radius $r_{\rm in}= 0.232$~au to an outer disk radius of $r_{\rm out}=300$~au.
The above value of $r_{\rm in}$ is set by being the radius of the DZIB in the $\dot{m}=10^{-9}\:M_\odot\:{\rm yr}^{-1}$ case, i.e., where $T$ reaches $1200 \:{\rm K}$. While the DZIB in higher accretion rate cases is at moderately larger radii, i.e., about 0.7~AU, for simplicity in all our models we follow the evolution down to 0.232~AU, recognising that this extends slightly inside the DZIB in the higher accretion rate cases: these very inner regions should be ignored as the disk structure here does not include the higher viscosity MRI-active region. 

The choice of the outer disk radius is somewhat arbitrary. Our approach is to extend to relatively large disk sizes that may be relevant to some observed systems, while recognising that most disks may be much smaller in reality (although still fed from infall from a larger-scale envelope). Thus, as discussed in \S\ref{subsec:cav} and Appendix~\ref{sec:appendix_1Myr}, for some cases the enclosed mass of the disk extending out to 300~au can be unrealistically large, i.e., the case with an initial accretion rate of $10^{-8}\:M_{\odot}\:{\rm yr}^{-1}$. In this particular case one should focus mainly on the results for the inner regions of the disk.

Figure \ref{fig:param} shows the radial profiles of $\Sigma$, $T$, $\kappa$ and $n_{\rm H}$ for disk models with $\dot{m}$ from $10^{-8}$ to $10^{-9}\:M_\odot\:{\rm yr}^{-1}$. Evolution along this sequence is the final fiducial model we will present. We note that the temperatures range from just over 1000~K in the inner regions down to below 20~K at 300~au. The number densities range from about $n_{\rm H}\sim 10^{16}\:{\rm cm}^{-3}$ in the inner regions down to $\sim 10^9\:{\rm cm}^{-3}$ at 300~au. 

We adopt a standard refractory dust-to-gas mass ratio of $0.7\%$ \citep[e.g.,][]{draine2011}, which we will use for the composition of material supplied to the outer disk and also, in some cases, for initial conditions within our disk models. We will also study models that start with pre-evolved compositions that can change due to grain/pebble drift, described below.

Note, from here on we will use the term grain and pebble interchangeably, i.e., pebbles can be considered to simply be large grains. The size of grains affects the rates of grain surface chemical reactions, gas-grain interactions, and freezeout/sublimation. Our starting models assume the midplane is populated by single size spherical grains/pebbles of radius $a_p$ and mass $m_p$. Typical models of interstellar molecular clouds \citep[e.g.,][]{entekhabi2021astrochemical} assume $a_p = 0.1 \mu$m and that the grains have bulk density $\rho_{p} = 3\:{\rm g\:cm}^{-3}$. For our model, this results in there being a grain number density of $n_p/n_{\rm H} = 1.30 \times 10^{-12}$ and a total grain area per H of $1.63 \times 10^{-21}\:{\rm cm}^2$. Models that are initialized with an increased grain size have a correspondingly reduced grain abundance to keep the initial local dust-to-gas mass ratio fixed. There is an impact on the chemistry via the reduced grain surface area per H-nuclei and an impact on the dynamics via changes to the radial 
drift
speeds, discussed in the next section. We will consider models with $a_p$ as large as $\sim$cm sizes.

\begin{figure}
    \centering
    \includegraphics[width=0.45\textwidth]{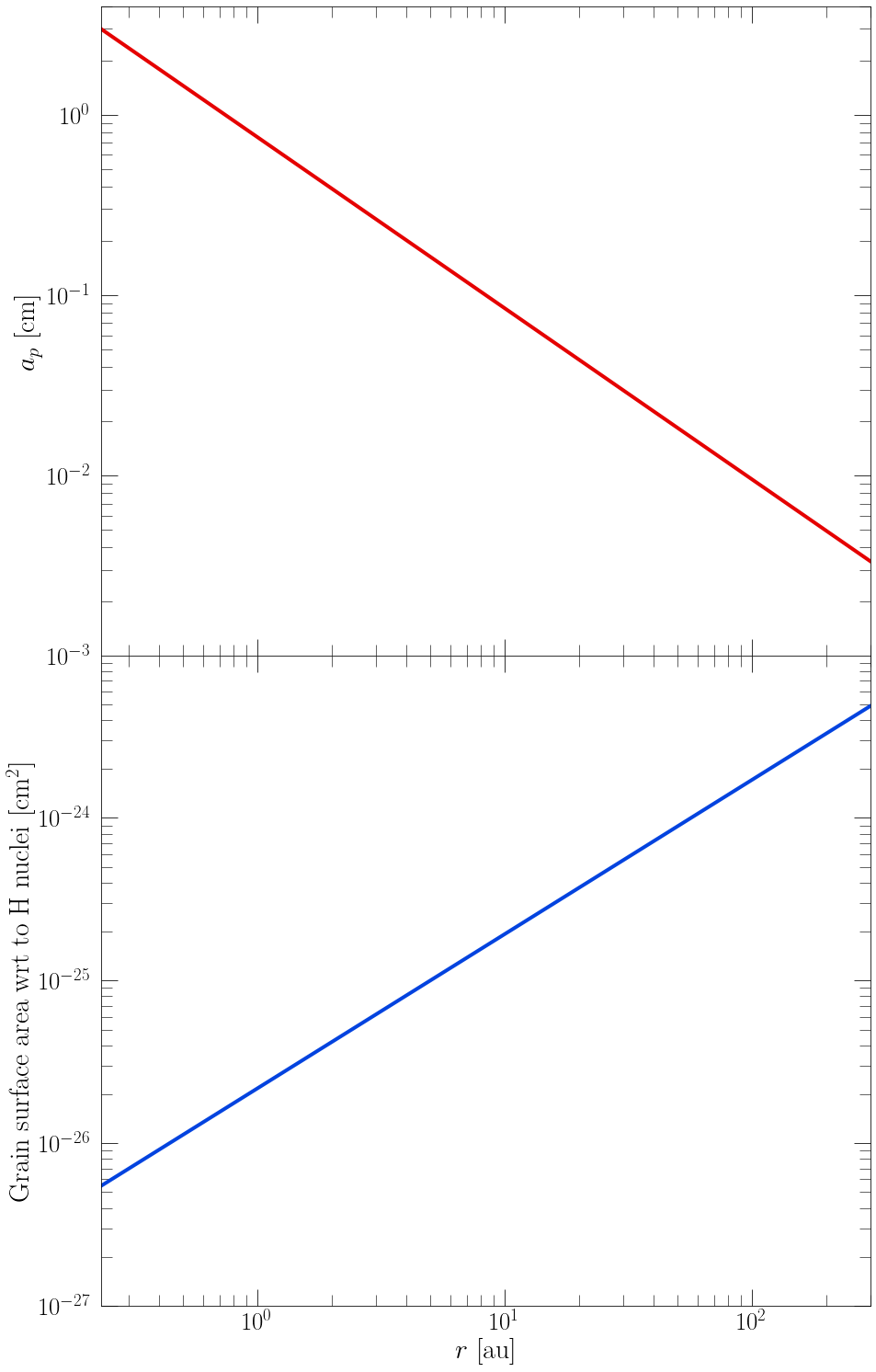}
    \caption{(a) Top: Adopted midplane grain/pebble size for Case (f) (described in \S\ref{sec:res}) based on the results of \citet{Hu_2018} (see eq.~\ref{eq:peb_profile}). (b) Bottom: Corresponding grain surface area per H nucleus versus radius.
    }
    \label{fig:grain_gradient}
\end{figure}

In addition to the single size models, we will also consider cases in which there is a radial gradient of $a_p$ within the disk, i.e., increasing from small values in the outer disk to larger values in the inner regions \citep[see, e.g.,][]{Birnsteil2012,Hu_2018}. Guided by such studies of pebble growth, we will describe $a_p(r)$ via: 
\begin{equation}\label{eq:peb_profile}
    \log_{10}{(a_p / {\rm cm})} = -0.9484 \log_{10}{(r_{\rm au})} - 0.1299,
\end{equation}
which is a profile with pebbles having radii of $\sim 3\:$cm at the DZIB and $\sim 3 \times 10^{-3}\:$cm at the outer boundary of the disk. This size variation and the corresponding evolution of the grain surface area with respect to H nuclei are shown in Figure \ref{fig:grain_gradient}. Note that in this model the surface area per H for gas-grain reactions, including the freeze-out from gas to ice phases, decreases significantly as one moves from the outer to inner disk.


\subsection{Gas and pebble radial velocities}

\begin{figure*}
    \centering
    \includegraphics[width=1.0\textwidth]{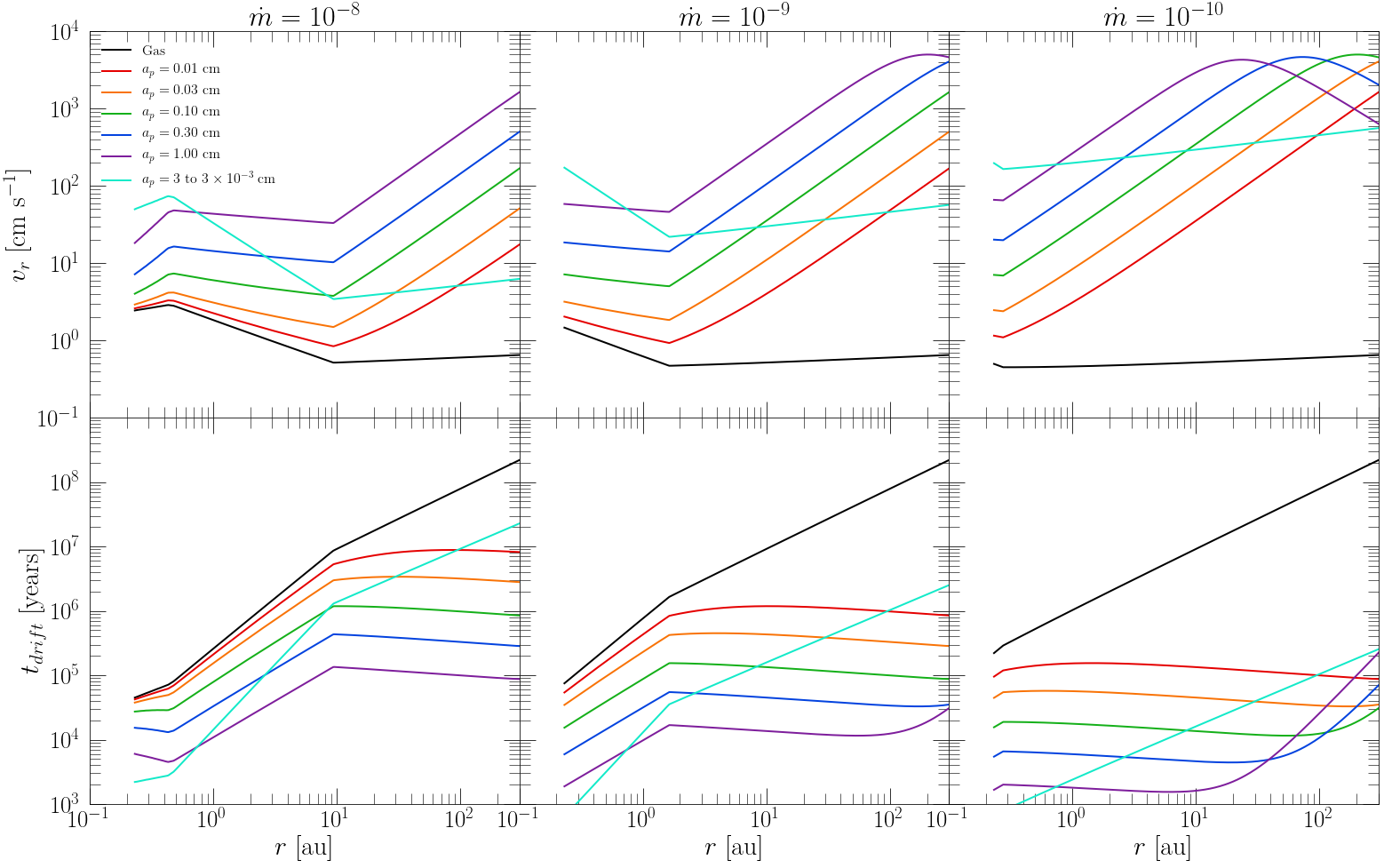}
    \caption{{\it (a) Top row:} Radial velocity profiles of pebbles in example disk models with accretion rates $\dot{m}=10^{-8}\:M_\odot\:{\rm yr}^{-1}$ (left), $10^{-9}\:M_\odot\:{\rm yr}^{-1}$ (middle), and $10^{-10}\:M_\odot\:{\rm yr}^{-1}$ (right). Results for single pebble size models from 0.01~cm to 1~cm are shown, as well the case of a radially evolving pebble size model from 0.003~cm in the outer disk to 3~cm in the inner disk (see legend). The radial speed of the gas is also indicated.
    {\it (b) Bottom row:} Radial drift times, $t_{\rm drift}\equiv r/v_{r}$ are shown for the same models shown in (a).}
    \label{fig:velocities}
\end{figure*}

Both gas and grains in the disk midplane have inward radial velocities towards the central star. The gas velocity, $v_{r,g}$, is given by:
\begin{equation}
    v_{r,g} = \frac{3 \alpha}{2} \frac{c_s}{v_K} c_s.
\end{equation}
The radial profiles of $v_{r,g}$ for various disk models are shown in Fig.~\ref{fig:velocities}, with typical values of $\sim 1\:{\rm cm\:s}^{-1}$.

The pebble drift arises due to gas drag. Small pebbles, e.g., with sizes of the order of micrometers, are very well-coupled to the gas and move with it on slightly sub-Keplerian orbits, given a typical gas pressure profile that decreases with radius, that spiral towards the central star. Larger pebbles orbit nearer to the Keplerian velocity and so experience a headwind of speed $v_\Delta$ that saps their angular momentum and causes them to spiral inwards at a faster rate than the gas. A frictional timescale can be defined via $t_{\rm fric}\equiv m_p v_\Delta / |F_D|$, where $F_D$ is the drag force. We follow the methods of \cite{Hu_2018} to treat the dynamics of the pebbles under the influence of gas drag, with the Epstein drag regime being the relevant case over most of the radii of our disk models, transitioning to the Stokes regime only in the very inner regions.

The radial drift speed of a pebble relative to the gas is
\begin{equation}
\label{eq:vrp}
    v_{r,p} \simeq -k_P (c_s/v_K)^2 (\tau_{\rm fric} + \tau_{\rm fric}^{-1})^{-1} v_K,
\end{equation}
where $k_P\simeq 2.4$
is the power-law index of pressure radius relation in $P=P_0 (r/r_0)^{-k_P}$, $\tau_{\rm fric} \equiv \Omega_K t_{\rm fric}$ is the dimensionless friction timescale and $\Omega_K \equiv v_K/r$ is the orbital angular frequency \citep{armitage_2009}.

Figure \ref{fig:velocities} shows the radial drift velocities obtained using equation \ref{eq:vrp} of various pebble sizes for the physical conditions of our model. This figure also shows the associated drift timescales, defined as $t_{\rm drift}\equiv r/v_{r,p}$. The equivalent velocities and timescales for gas advection are also shown. As the pebble size increases, so does the drift velocity. The lower accretion rate models, with lower gas density and temperature profiles, tend to have faster drift speeds.
In particular, for the $\dot{m}=10^{-8}\:M_{\odot}\:{\rm yr}^{-1}$ accretion rate models, the drift speeds of pebbles reveal that only those initially situated $\lesssim10$~au are able to enrich the innermost regions of the disk within $10^5$ years. Similarly, this applies to only that gas at $<1$~au. Shorter drift times for the $\dot{m}=10^{-9}\:M_{\odot}\:{\rm yr}^{-1}$ accretion rate model allows pebbles from further out to influence the inner disk.

\begin{table*}
\caption{Initial/Boundary abundances with respect to hydrogen nuclei for atomic, molecular, and pre-aged abundance scenarios with three different CRIRs. The abundance of grains is for the case with 
grain radius $a_p = 10^{-5}$~cm. Pre-aged abundances shown here are based on a model adopting this grain size and abundance.
\label{tab:initial}}
\begin{tabular}{lllllllll}
\hline
Species & Atomic & Molecular & Pre-aged (Gas) & Pre-aged (Ice) & Pre-aged (Gas) & Pre-aged (Ice) & Pre-aged (Gas) & Pre-aged (Ice)\\

 & & & CRIR-16 & CRIR-16 & CRIR-17 & CRIR-17 & CRIR-18 & CRIR-18\\
\hline
H$_2$ & $5.000 \times 10^{-01}$ & $5.000 \times 10^{-01}$ & $5.000 \times 10^{-01}$ & $1.048 \times 10^{-06}$ & $5.000 \times 10^{-01}$ & $1.076 \times 10^{-05}$ & $5.000 \times 10^{-01}$ & $1.041 \times 10^{-06}$\\
H & $5.000 \times 10^{-05}$ & $5.000 \times 10^{-05}$ & $1.183 \times 10^{-08}$ & $5.407 \times 10^{-16}$ & $1.260 \times 10^{-12}$ & $9.477 \times 10^{-24}$ & $1.257 \times 10^{-10}$ & $7.317 \times 10^{-25}$\\
He & $9.750 \times 10^{-02}$ & $9.800 \times 10^{-02}$ & $9.750 \times 10^{-02}$ & $3.343 \times 10^{-17}$ & $9.750 \times 10^{-02}$ & $3.434 \times 10^{-16}$ & $9.750 \times 10^{-02}$ & $3.323 \times 10^{-17}$\\
O & $1.800 \times 10^{-04}$ &  & $2.624 \times 10^{-10}$ & $4.317 \times 10^{-19}$ & $7.791 \times 10^{-13}$ & $7.009 \times 10^{-20}$ & $3.264 \times 10^{-12}$         & $7.208 \times 10^{-20}$           \\
N & $7.500 \times 10^{-05}$ &  & $1.761 \times 10^{-10}$         & $9.831 \times 10^{-15}$            & $7.370 \times 10^{-13}$ & $6.700 \times 10^{-14}$ & $3.042 \times 10^{-12}$ & $2.622 \times 10^{-14}$        \\
S & $8.000 \times 10^{-08}$ &  & $1.177 \times 10^{-13}$         & $2.190 \times 10^{-23}$            & $1.667 \times 10^{-16}$ & $4.182 \times 10^{-23}$ & $8.128 \times 10^{-16}$ & $3.090 \times 10^{-24}$        \\
F & $2.000 \times 10^{-08}$ &  & $1.150 \times 10^{-08}$         & $7.185 \times 10^{-14}$           & $2.000 \times 10^{-08}$ & $1.723 \times 10^{-12}$ & $2.000 \times 10^{-08}$ & $1.668 \times 10^{-13}$     \\
Si & $8.000 \times 10^{-09}$ &  & $1.191 \times 10^{-14}$        & $4.929 \times 10^{-18}$           & $5.344 \times 10^{-18}$ & $1.449 \times 10^{-11}$ & $1.619 \times 10^{-17}$ & $8.155 \times 10^{-12}$  \\
Cl & $4.000 \times 10^{-09}$ &  & $3.379 \times 10^{-15}$        & $1.473 \times 10^{-17}$           & $2.414 \times 10^{-16}$ & $1.438 \times 10^{-09}$ & $2.835 \times 10^{-16}$        & $1.875 \times 10^{-09}$ \\
P & $3.000 \times 10^{-09}$ &  & $2.723 \times 10^{-13}$     & $2.945 \times 10^{-09}$           & $2.868 \times 10^{-14}$ & $2.962 \times 10^{-09}$ & $2.963 \times 10^{-13}$        & $2.962 \times 10^{-09}$\\
$\rm H_2O$ & & $3.000 \times 10^{-04}$ & $6.374 \times 10^{-12}$         & $1.499 \times 10^{-05}$           & $5.317 \times 10^{-14}$ & $1.125 \times 10^{-05}$ & $3.036 \times 10^{-14}$ & $1.131 \times 10^{-05}$\\
$\rm H_2O_2$ & & & $3.622 \times 10^{-11}$         & $2.808 \times 10^{-06}$           & $1.122 \times 10^{-13}$ & $1.756 \times 10^{-06}$ & $1.125 \times 10^{-14}$        & $1.763 \times 10^{-06}$           \\
CO & $1.400 \times 10^{-04}$ & $6.000 \times 10^{-05}$ & $2.262 \times 10^{-10}$ & $1.394 \times 10^{-04}$ & $3.002 \times 10^{-12}$ & $1.395 \times 10^{-04}$ & $1.142 \times 10^{-11}$ & $1.399 \times 10^{-04}$\\
$\rm CO_2$ & & $6.000 \times 10^{-05}$ & $1.361 \times 10^{-13}$         & $5.756 \times 10^{-07}$           & $7.391 \times 10^{-15}$ & $4.865 \times 10^{-07}$ & $7.305 \times 10^{-16}$ & $5.618 \times 10^{-08}$\\
$\rm CH_4$ & & $1.800 \times 10^{-05}$ & $9.797 \times 10^{-15}$         & $8.067 \times 10^{-09}$           & $5.703 \times 10^{-18}$ & $9.407 \times 10^{-11}$ & $1.138 \times 10^{-17}$ & $9.448 \times 10^{-11}$\\
$\rm N_2$ & & $2.100 \times 10^{-05}$ & $2.171 \times 10^{-10}$        & $3.475 \times 10^{-05}$            & $1.845 \times 10^{-11}$ & $3.070 \times 10^{-05}$ & $1.931 \times 10^{-10}$ & $3.116 \times 10^{-05}$\\
$\rm NH_3$ & & $2.100 \times 10^{-05}$ & $8.700 \times 10^{-12}$        & $4.350 \times 10^{-07}$           & $3.073 \times 10^{-15}$ & $2.947 \times 10^{-07}$ & $1.890 \times 10^{-14}$ & $2.752 \times 10^{-07}$\\
$\rm H_2S$ & & $6.000 \times 10^{-06}$ & $2.890 \times 10^{-15}$        & $7.844 \times 10^{-09}$           & $2.160 \times 10^{-17}$ & $1.080 \times 10^{-09}$ & $3.690 \times 10^{-17}$ & $1.439 \times 10^{-09}$\\
$\rm O_2$ &  &  & $1.218 \times 10^{-11}$        & $7.688 \times 10^{-05}$           & $4.458 \times 10^{-13}$ & $7.570 \times 10^{-05}$ & $4.174 \times 10^{-13}$        & $7.635 \times 10^{-05}$           \\
HCN &  &  & $1.175 \times 10^{-16}$        & $2.656 \times 10^{-10}$            & $1.451 \times 10^{-16}$ & $2.181 \times 10^{-10}$ & $4.615 \times 10^{-17}$        & $6.588 \times 10^{-12}$           \\
HNC &  &  & $3.792 \times 10^{-16}$        & $1.157 \times 10^{-10}$            & $2.495 \times 10^{-17}$ & $1.725 \times 10^{-12}$ & $1.568 \times 10^{-17}$        & $1.612 \times 10^{-12}$           \\
NO &  &  & $1.640 \times 10^{-13}$        & $5.439 \times 10^{-13}$           & $4.146 \times 10^{-13}$ & $8.244 \times 10^{-06}$ & $1.144 \times 10^{-13}$        & $1.207 \times 10^{-05}$        \\
HNO &  &  & $9.827 \times 10^{-12}$        & $5.057 \times 10^{-06}$           & $8.556 \times 10^{-12}$ & $3.311 \times 10^{-06}$ & $3.653 \times 10^{-13}$     & $3.230 \times 10^{-07}$           \\
Grains & $1.300 \times 10^{-12}$ & $1.300 \times 10^{-12}$ &         & $1.300 \times 10^{-12}$           &  & $1.300 \times 10^{-12}$ &         & $1.300 \times 10^{-12}$           \\
\end{tabular}
\end{table*}

\subsection{Chemical model}

We make use of the time-dependent gas-grain astrochemical network developed by \cite{Walsh_2015} to compute the chemical evolution in each radial zone in the disk. The network consists of 8764 reactions that evolve the abundances of 668 species in gas and ice-mantle phases. The network was extracted from the UMIST 2012 Database for Astrochemistry \citep[see][]{McElroy_2013} and then complemented by both thermal and non-thermal gas-grain interactions, including cosmic-ray and X-ray induced desorption, photodesorption, and grain-surface reactions. An overview of the chemical network and the reaction types included are described in detail in \cite{Walsh_2010} and \cite{Walsh_2015}.

For our model, we set the cosmic ray ionization rate (CRIR) at a constant value of $\zeta_{cr} = 1.0 \times 10^{-17}$ s$^{-1}$ for all positions in the disk. For simplicity, we choose to ignore any attenuation effects that may arise in the high mass surface density inner regions of the disk \citep[see, e.g,][]{umebashy1981}.  Also, we do not include any X-ray background ($\zeta_{xr} = 0$ s$^{-1}$), which we consider to be a reasonable first approximation for disk midplane conditions. Furthermore, also for simplicity, we do not include effects of cosmic-ray induced thermal desorption, especially given the inherent uncertainties about the process and the expectation it will become a weaker effect for larger grains that suffer smaller induced thermal fluctuations. Given our focus on midplane conditions, we set $A_V=100$ mag, which means that there is negligible influence of any reasonable background FUV radiation field.

For the initial chemical composition of the disk and the material supplied at its outer boundary we consider three different scenarios: atomic, molecular, and pre-aged abundances (see Table \ref{tab:initial}). Note, quoted abundances are always with respect to the total number of H-nuclei. Both atomic and molecular cases assume all species are found in gas form, with the grains still vacant of ice.

The atomic scenario assumes a chemical ``reset'', where the material from the molecular infalling envelope suffers significant chemical changes due to shock heating as it reaches the disk.
This mimics early chemical models where the initial conditions were assumed to be atomic \citep[see, e.g.,][]{Willacy_1998}.
The molecular scenario is a simple method of investigating another scenario where molecular conditions in the gas phase are adopted. This is also known as the ``inheritance'' scenario in which it is assumed that comet/interstellar ice abundances are inherited by the protoplanetary disk \citep[see, e.g.,][]{Marboeuf_2014}.

For the pre-aged scenario, which we consider to be most self-consistent and perhaps the most realistic, we start with atomic species and fiducial-size, small dust grains and then chemically evolve for $10^{4}$ years at conditions found at $300$~au, i.e., the outer boundary of our disk. In this region, the density and temperature reach down to values of $n_{\rm H} \sim 10^9 \:{\rm cm}^{-3}$ and $T \sim 15 \:{\rm K}$. Given the relatively high densities, we find that the resulting abundances do not change significantly after this timescale and also do not depend appreciably on the choice of atomic initial abundances.
This chemical pre-aging results in virtually all carbon atoms and most oxygen atoms being locked in CO ice. Similarly, most nitrogen atoms end up as $\rm N_2$ ice.

\subsection{Combined chemical evolution and advective transport}

We model the disk as a series of discrete cells spaced evenly in ${\rm log}\:r$. Each cell, representing an annular 3D ring, is assumed to have constant physical properties, e.g., gas density and temperature, that are set by the expected values from the $\alpha$-disk model at the center of the cell. Cells are assumed to have a vertical extent above the midplane equal to the scaleheight evaluated at cell centre. This then gives each cell a defined volume and mass.

The coupling of the chemical evolution and advective transport of material is done in two separate steps. First, chemical evolution from $t$ to $t+dt$ is computed using the astrochemical network for each cell given its density and temperature.

Second, for the advective transport step we compute the fluxes between cells using the velocities of gas and pebbles estimated at each cell boundary. The flux of gas is normalized to yield a constant accretion rate at all radial locations in the disk, which in practice means that the cell vertical height at its inner boundary is adjusted by a small dimensionless factor given that the viscous accretion rate depends on some quantities evaluated at cell centre. The advective abundances are assumed to be averages of conditions before and after the current chemical evolution step. The material that is advected into a cell is assumed to mix immediately and completely with that already present. Abundances of chemical species and dust grains / pebbles
are then updated for each cell.

Note that in our models all the advection is radially inwards. At the inner boundary, material is simply advected out of the simulation domain, i.e., into the innermost disk regions that are not included in our set up. At the outer cell, material flows in given the assumptions of the composition of the boundary material, i.e., as described in Table \ref{tab:initial}.

In most of our models we have a fixed pebble size independent of radius. However, in the case where pebble size has a defined radial profile, the advective pebble material is assumed to immediately achieve the local desired size on reaching the next inward cell.

The time step, $dt$, of the calculation is limited by preventing the advected distance to be a percentage, 90\%, of the cell radial thickness. In practice, this is set by pebble radial drift speeds across the innermost zones.

\section{Results}\label{sec:res}

This section presents the results from our models for the coupled physical and chemical evolution of key volatile species that serve as the main reservoirs for C, O and N, i.e., CO, $\rm CO_2$, $\rm O_2$, $\rm CH_4$, $\rm H_2O$, $\rm H_2O_2$, $\rm N_2$, NO, HNO, HCN, HNC and $\rm NH_3$. However, since the chemical evolution also depends on dust/pebble mass to gas mass ratio, we need to understand its behavior alongside that of the species abundances. Figure \ref{fig:dustratio} presents the final state of this dust-to-gas mass ratio, while Figures \ref{fig:abu} (for main C and O-bearing species) and \ref{fig:nitro} (for main N-bearing species) present the relative chemical abundances as a function of radial distance from the central star. In each figure we present a sequence of models that gradually add realism and build toward our final, fiducial model. Unless otherwise specified, all models utilise pre-aged initial and boundary conditions and are then evolved for $10^5\:$yr.

\textbf{Case (a)} is a disk with a fixed accretion rate of $\dot{m} = 10^{-9}\:M_{\odot}\:{\rm yr}^{-1}$, dust/pebbles of size $a_p=10^{-5}$~cm, and all transport mechanisms turned off. \textbf{Case (b)} is the same as (a), but increases the pebble size to $a_p = 0.3$~cm. \textbf{Case (c)} is the same as (b), but includes gas radial advection and assumes the pebbles move with the gas. \textbf{Case (d)} is the same as (c), but decouples the pebbles from the gas, i.e., applying the appropriate pebble drift velocity. \textbf{Case (e)} is the same as (d), but involves an evolution of the accretion rate, starting from $\dot{m} = 10^{-8}\:M_\odot\:{\rm yr}^{-1}$ and declining to $\dot{m} = 10^{-9}\:M_{\odot}\:{\rm yr}^{-1}$ within $10^5\:$yr using equation \ref{eq:mdot}. Finally, \textbf{Case (f)} is the same as (e), but assumes a pebble size profile with radius in the disk, as described by eq.~\ref{eq:peb_profile} and shown in Fig.~\ref{fig:grain_gradient}. A summary of the information about Cases (a) to (f) is given in Table \ref{tab:cases}.

\begin{table*}
\caption{Summary of physical properties for the six cases being investigated in this work.
\label{tab:cases}}
\begin{tabular}{lllll}
\hline
Case & Transport Type & Mass accretion rate ($\dot{m}$) & Pebble radius ($a_p$)\\
 & & ($M_\odot\:{\rm yr}^{-1}$)& (${\rm cm}$)\\
\hline
(a) & No transport & $10^{-8}$ & $10^{-5}$\\
(b) & No transport & $10^{-8}$ & $0.30$\\
(c) & Gas advection & $10^{-8}$ & $0.30$\\
(d) & Gas advection + Pebble drift & $10^{-8}$ & $0.30$\\
(e) & Gas advection + Pebble drift & $10^{-8} \rightarrow 10^{-9}$ & $0.30$\\
(f) & Gas advection + Pebble drift & $10^{-8} \rightarrow 10^{-9}$ & $3.00 \rightarrow 3 \times 10^{-3}$\\

\end{tabular}
\end{table*}

\subsection{Dust-to-gas mass ratio}

\begin{figure*}
    \centering
    \includegraphics[width=0.9\textwidth]{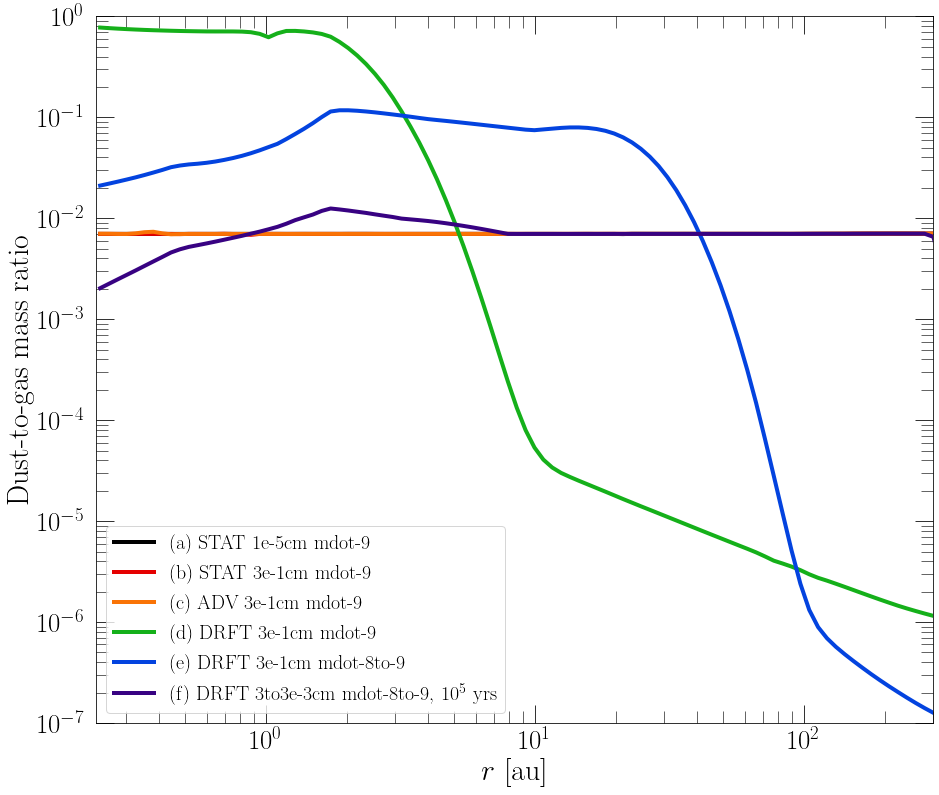}
    \caption{Dust-to-gas mass ratio (D/G) after $10^5$ years of evolution. The model sequence from Case (a) to (f) is described in the text. The legend also describes their key parameters. The first term is transport type: STAT is for no transport (static); ADV is for gas advection (used for both gas and grains); and DRFT is for gas advection plus radial drift of pebbles. The second term denotes grain size. The third term states the accretion rate condition. Note, Cases (a)---(c) all have the same, radially constant value of D/G.}
    \label{fig:dustratio}
\end{figure*}{}

Figure \ref{fig:dustratio} shows the values of the dust-to-gas mass ratio (D/G) of every case after $10^5$ years of evolution.
Cases (a)---(c) produce an unchanging D/G ratio at all disk radii due to these models featuring no transport or having equal mass fluxes for gas and dust components. 
However, as seen in Cases (d)---(f), when pebble drift is turned on the decoupling between gas and pebble velocity provokes a general depletion of the dust component in the outer disk. Note that while new dust is being added to the outer zone with its rate set in proportion to the gas accretion rate, this rate of supply is much lower than the rate of inward drift once in the disk. Thus the D/G ratio steadily decreases until an equilibrium level is reached. For example, in Case (d) the D/G ratio stabilizes at about $10^{-6}$ in the very outer disk at 300~au, rising to several $\times 10^{-5}$ by 10~au. Interior to this, the D/G ratio is still evolving at $10^5\:$yr, influenced by the initial condition value of 0.007 as well as the transient wave of inward moving pebbles delivered from the outer disk. Indeed, this wave of pebbles leads to D/G values that can be very large, i.e., approaching unity inside 2~au in the Case (d) disk.

In Case (e), with the introduction of an evolving accretion rate, the relatively lower radial velocities of pebbles when compared to the gas velocity (see Figure \ref{fig:velocities}) result in a slower evolution of the D/G ratio. The introduction of the pebble size profile in Case (f) slows the evolution of the dust component to a crawl in the outer regions, while the inner regions become those having the most noticeable D/G evolution, including only a modest enhancement from the inward wave and an inner region of dust depletion compared the initial condition value.

\begin{figure*}
    \centering
    \includegraphics[width=1.0\textwidth]{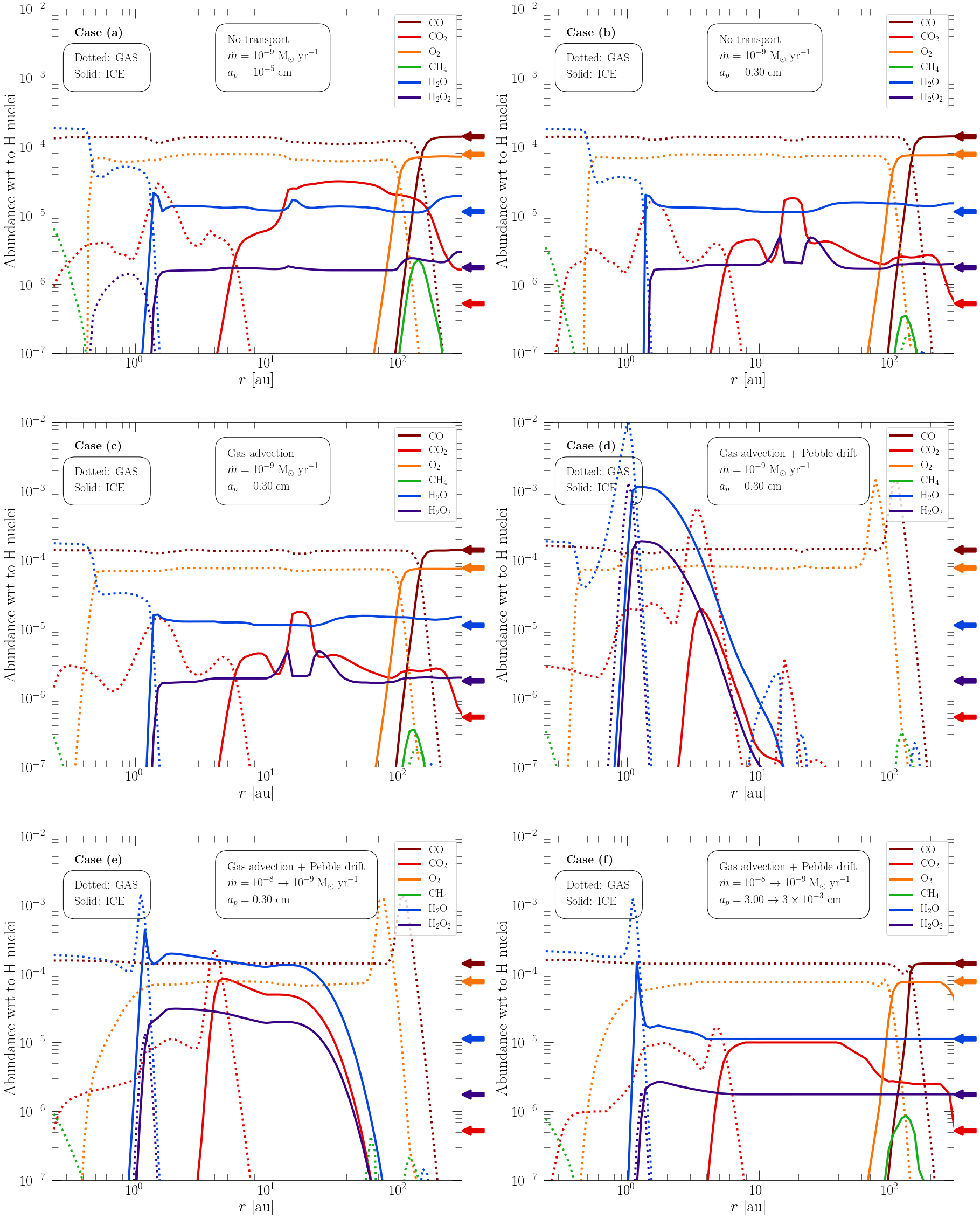}
    \caption{Final abundances of C and O-bearing species with respect to H nuclei after $10^5$ years of evolution of the disk for cases (a) to (f) (see figure legends and main text). Dotted lines represent gas species, while solid lines are for ice species. Arrows to the right of each plot mark the initial ice-phase abundances of each species as listed in Table \ref{tab:initial}.
    }
    \label{fig:abu}
\end{figure*}{}

\begin{figure*}
    \centering
    \includegraphics[width=1.0\textwidth]{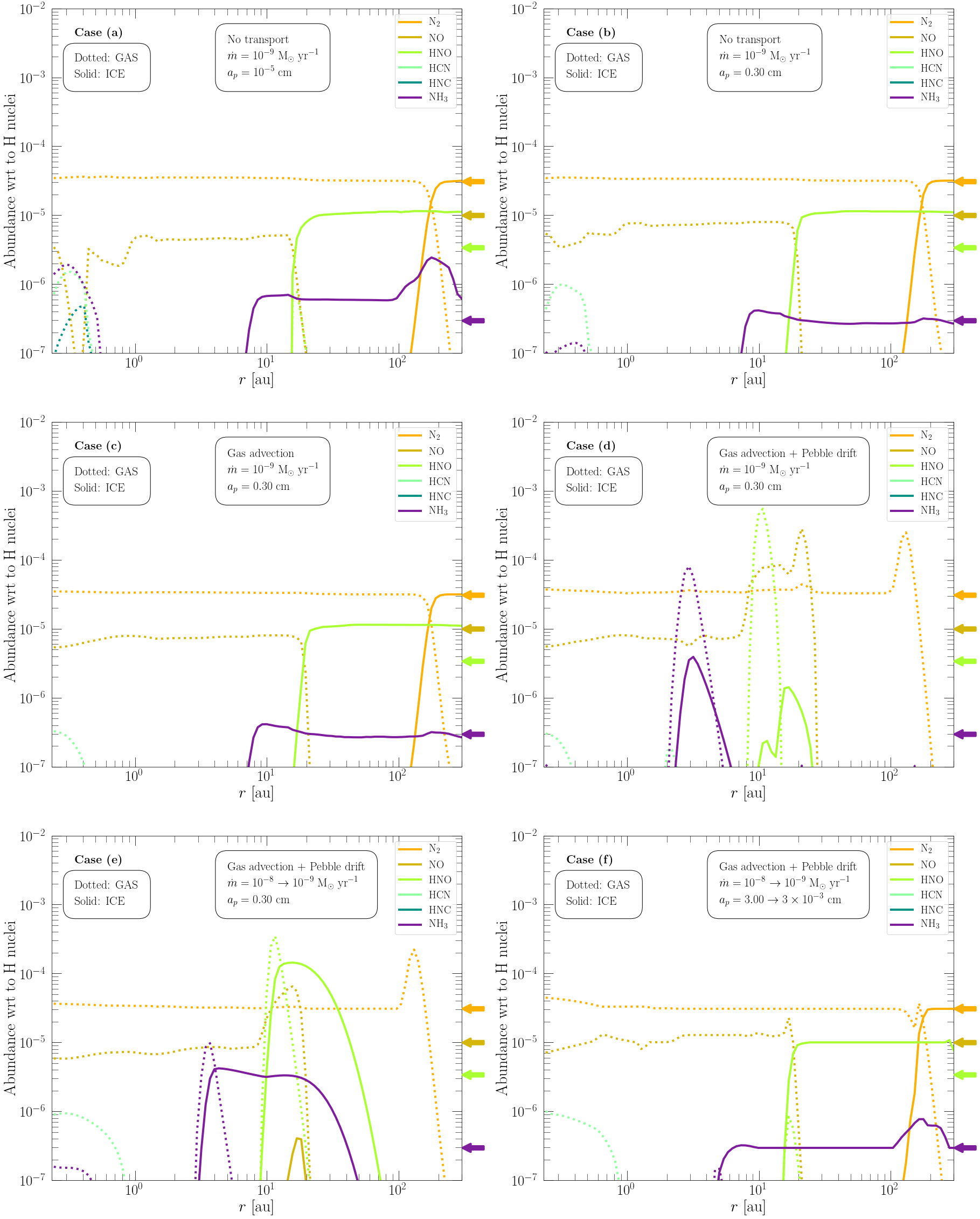}
    \caption{As Figure~\ref{fig:abu}, but for N-bearing species.}
    \label{fig:nitro}
\end{figure*}{}

\subsection{Chemical evolution of static cases}

Figures \ref{fig:abu}a and \ref{fig:abu}b (for main C and O-bearing species) and Figures~\ref{fig:nitro}a and \ref{fig:nitro}b (from main N-bearing species) show the final abundances of key volatile species after $10^5$ years of evolution of the disk for models where only chemical evolution is active, i.e., transport mechanisms are turned off and all radially dependant physical parameters are held fixed. These models only differ in the size of the pebbles, with Case (a) having $a_p = 10^{-5}\:$cm and Case (b) having the much larger $a_p = 0.30\:$cm.
In both cases, CO and $\rm O_2$ start as ices with abundances with respect to H nuclei of $1.40 \times 10^{-4}$ and $7.57 \times 10^{-5}$, respectively, making CO ice the main reservoir of C, while CO ice and $\rm O_2$ ice are together the main reservoirs of O. We note that the fiducial initial $\rm H_2O$ ice abundance is $1.12\times 10^{-5}$.
$\rm N_2$ starts with an abundance of $3.07 \times 10^{-5}$, making it the main reservoir of nitrogen. With the presence of a radial temperature gradient, these ice species enter the gas phase at particular locations, i.e., ``icelines''.


In these models, the radial evolution of the main C-bearing species is dominated by CO and its transition from the ice phase in the outer disk, i.e., below $\sim 20\:$K, to the gas phase in the warmer inner regions. This iceline is located at about 140~au. The CO also carries a significant part, about 40\%, of the O reservoir, with the remainder mostly contained in $\rm O_2$, which has its iceline just interior to 100~au. We see that $\rm H_2O$ ice remains at a fairly constant level, just above an abundance of $10^{-5}$, until reaching its iceline at about 1.3~au. However, a small part of the water ice is hydrogenated to form $\rm H_2O_2$ \citep[see, e.g.,][]{Semenov2011}. In the inner $\sim 1\:$au, gas phase water becomes a more important carrier for O, eventually dominating in the innermost regions at the expense of gas phase $\rm O_2$. This is caused by a new chemical pathway opening up above $\sim 1,000\:$K permitting gas-phase $\rm O_2$ to react with H atoms to form OH, that in turn produces $\rm H_2O$ gas \citep[see, e.g.,][]{Walsh_2015}. 


The next most important species is $\rm CO_2$. We see that $\rm CO_2$ ice shows a significant increase from its initial abundance of $4.87 \times 10^{-7}$ to about $10^{-5}$ as one moves inwards from about 100~au to 10~au. As described by \cite{Eistrup_2016} and \cite{Eistrup_2018}, and explained by \cite{Walsh_2015}, ice-phase $\rm CO_2$ abundance grows at temperatures $>20\:$K from gas-phase CO molecules making contact with OH radicals produced by CR-induced UV photodissociation of $\rm H_2O$ ice. This produces the $\rm CO_2$ ice {\it in situ} on the surface of the grains. The $\rm CO_2$ iceline is reached at about 5~au.
Interior to this, the $\rm CO_2$ gas phase abundance fluctuates by factors of several, especially being enhanced near the water iceline and then decreasing in the very innermost regions, where $\rm CH_4$ newly arises. However, CO and $\rm H_2O$ are by far the most dominant C and O-bearing species in this inner zone.


For the N-bearing species, $\rm N_2$ remains the main carrier at all radii in the Case (a) and (b) models, with the iceline located just inside 200~au. The next most important species are NO and HNO ices \citep[see, e.g.,][]{Semenov2011}. They are present with abundances a few times lower than that of $\rm N_2$ down to their iceline radii of about 15~au. Next in importance in the ice phase is $\rm NH_3$ with an abundance in the range of about $10^{-7}$ to $10^{-6}$ and with its iceline at about 7~au. 

With dust mass being a fixed quantity, increasing the size of the grains reduces the surface area and thus reaction rates of surface reactions.
As discussed by \cite{Eistrup_2016} and \cite{Eistrup_2018}, an increased amount of grains would allow for greater collision rates between them and CO gas. Ice-phase CO then quickly reacts with OH before it desorbs, readily forming $\rm CO_2$ ice. In our cases, chemical pathways dependant on grain surfaces that lead to the creation of ice-phase $\rm CO_2$, $\rm CH_4$, and $\rm NH_3$, and gas-phase $\rm CO_2$, HNO, $\rm NH_3$, HCN and HNC are very visibly affected by a change in grain surface area per H.

To demonstrate the effects of varying the initial/boundary conditions, the CRIR, and to compare our results with those presented in \cite{Eistrup_2016} for their \textit{Full chemistry} models, Appendix \ref{sec:appendix} contains Figures \ref{fig:abu1_NT} and \ref{fig:abu2_NT} that show models featuring no transport but combinations of the initial/boundary conditions summarized in Table \ref{tab:initial} chemically evolved with different values for the CRIR. These exhibit the strong influence of initial conditions on the final composition of the disk after $10^5$ years, as well as the strong effect of the CRIR on the rate at which the dominant reservoirs like CO and $\rm N_2$ are destroyed and turned into other species. Higher levels of ionization rate tend to equalize disk compositions regardless of initial/boundary conditions. Differences between our results and those of \cite{Eistrup_2016} can be explained by slightly different choices in initial conditions and our shorter evolution time (their models run for 1 Myr).

\subsection{Chemical evolution of advective cases}

Figures \ref{fig:abu}c and \ref{fig:nitro}c show that the implementation of the gas radial velocity to both gas and solid components has little effect on the overall abundances compared to the previous cases with no transport (specifically in comparison to Figures~\ref{fig:abu}b and \ref{fig:nitro}b). The radial drift time for the gas velocity at this accretion rate indicates that noticeable changes would only appear after $10^5$ years in regions $\lesssim 1$~au. Thus, only the innermost cells have their components pushed inwardly with sufficiently large speeds that they manage to influence the final composition of their neighbors. This is clearly visible for gas-phase $\rm CO_2$, $\rm CH_4$, HCN, and $\rm NH_3$ at $\lesssim 0.3$ AU.


However, as shown in Figures \ref{fig:abu}d and \ref{fig:nitro}d, the effects of advection of the solid phases via radial pebble drift are much more dramatic.  Tied to the previously described evolution of the dust-to-gas mass ratio, after $10^5\:$yr the outer disk ($\gtrsim 10\:$au) abundances of the ices of CO, $\rm CO_2$, $\rm O_2$, $\rm H_2O$, $\rm CH_4$, $\rm N_2$ and $\rm NH_3$ become much lower, by several orders of magnitude, compared to the models without radial pebble drift. Inside about 10~au, remaining ice species, such as $\rm H_2O$, $\rm H_2O_2$, $\rm CO_2$ and $\rm NH_3$, show large enhancements. The sublimation of these species at their icelines also leads to increased gas phase abundances, especially since the gas is then assumed to be moving at much smaller radial inwards velocities. For example, we see gas phase $\rm H_2O$ peaking at an abundance of $\sim 10^{-2}$ with respect to H nuclei at 1~au. Similarly, gas phase $\rm NH_3$ reaches an abundance of $\sim 10^{-4}$ at about 3~au. Further inwards, the gas phase abundances decrease down to the levels previously seen in the models without pebble drift, since there has not been time for gas advection to spread the local enhancements all the way in from the icelines.

\subsection{Evolving accretion rate}

\begin{figure*}
    \centering
    \includegraphics[width=1.0\textwidth]{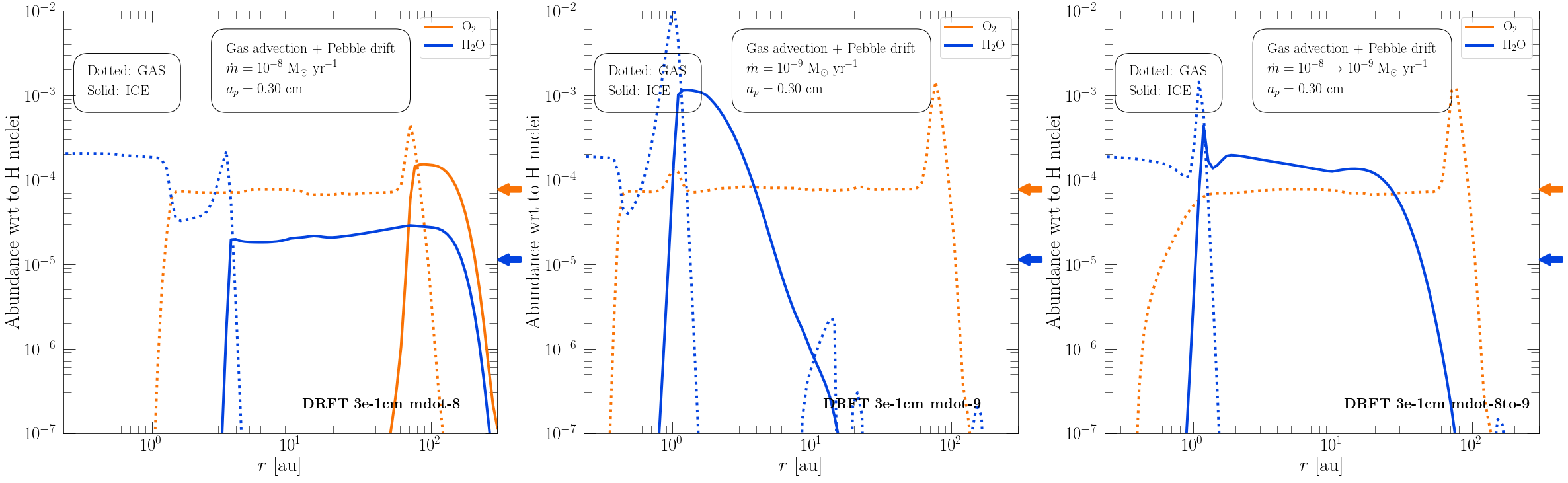}
    \caption{Three cases evidencing the varying location of icelines and chemistry for different accretion rates. \textit{(a) Left:} Disk with steady accretion rate of $\dot{m} = 10^{-8}\:M_{\odot}\:{\rm yr}^{-1}$. \textit{(b) Middle:} Disk with steady accretion rate of $\dot{m} = 10^{-9}\:M_{\odot}\:{\rm yr}^{-1}$. \textit{(c) Right:} Disk with evolving accretion rate varying from $\dot{m} = 10^{-8} \rightarrow 10^{-9}\:M_{\odot}\:{\rm yr}^{-1}$.}
    \label{fig:mdot-comp}
\end{figure*}

Figures \ref{fig:dustratio}, \ref{fig:abu}e and \ref{fig:nitro}e show the model with an exponentially declining accretion rate, starting from $\dot{m} = 10^{-8}\:M_{\odot}\:{\rm yr}^{-1}$ and ending with $\dot{m} = 10^{-9}\:M_{\odot}\:{\rm yr}^{-1}$. As pebble drift speeds are generally slower at greater gas densities (i.e., at higher accretion rates) in the spatial range of the models (see Figure \ref{fig:velocities}), the D/G ratio result of Case (e) can be interpreted, approximately, as being a less advanced state of dust evolution compared to that of Case (d). Thus, the region of depletion of dust is further out, i.e., beyond about 50~au. Interior to the this, the D/G is elevated by about a factor of 10. 

The abundance results for Case (e) demonstrate the migration of icelines and other localized regions of chemical processing within the active region of the disk. Note that the outer extent of the active region decreases from about 10~au to 2~au during the evolution of the Case (e) modelñ. 

Figure \ref{fig:mdot-comp} shows this behavior for $\rm O_2$ and $\rm H_2O$ by comparing models of constant accretion rates of $10^{-8}\:M_{\odot}\:{\rm yr}^{-1}$ and $10^{-9}\:M_{\odot}\:{\rm yr}^{-1}$ with the evolving model $\dot{m} = 10^{-8} \rightarrow 10^{-9}\:M_{\odot}\: {\rm yr}^{-1}$. We see that the region of high temperature where $\rm O_2$ is destroyed shrinks as the accretion rate drops. Similarly, the region of the $\rm H_2O$ iceline is drawn closer towards the star for lower accretion rates.

If we allow Case (e) to evolve for a longer time while keeping the final accretion rate $\dot{m} = 10^{-9}\:M_{\odot}\:{\rm yr}^{-1}$ constant then it would eventually appear similar to Case (d), though not exactly the same due to the impact of the decaying accretion rate on the initial 0.1 Myrs.

\subsection{Pebble size profile}

Figures \ref{fig:abu}f and \ref{fig:nitro}f, along with previously discussed Figure \ref{fig:dustratio}, demonstrate how the introduction of the pebble size dependence with disk radius, combined with the accretion rate evolution, decrease the effects of pebble drift on the D/G ratio and thus the variation of ice phase abundances. The outer disk now contains much smaller pebbles, which thus have near negligible drift rates. There are thus only very modest effects of slightly enhanced gas-phase $\rm O_2$, CO, $\rm N_2$, and NO after $10^5$ years.
Only at radii below 10 AU, where pebbles are assumed to be larger, is their drift time $t_{\text{drift}}$ low enough to produce a noticeable enhancement of gas-phase species, i.e., for $\rm H_2O$, $\rm H_2O_2$, $\rm CO_2$, and $\rm NH_3$.

Appendix \ref{sec:appendix_1Myr} shows a comparison between this fiducial Case (f) and a version, Case (g), which is evolved for 1 Myr, with the accretion rate also decaying in this interval instead of the original $10^{5}$ years. The longer timescale of evolution affects both the profiles of dust-to-gas ratio and chemical composition. In particular, abundances of gas phase $\rm H_2O$, $\rm CO_2$ and $\rm NH_3$ show enhancements in the regions just interior to their respective icelines. These enhancements are due to a combination of enhanced delivery of icy pebbles to these regions, as well as more time for gas advection interior to the ice line.

\subsection{Total C/H, O/H and N/H abundance ratios}

\begin{figure*}
    \centering
    \includegraphics[width=1.0\textwidth]{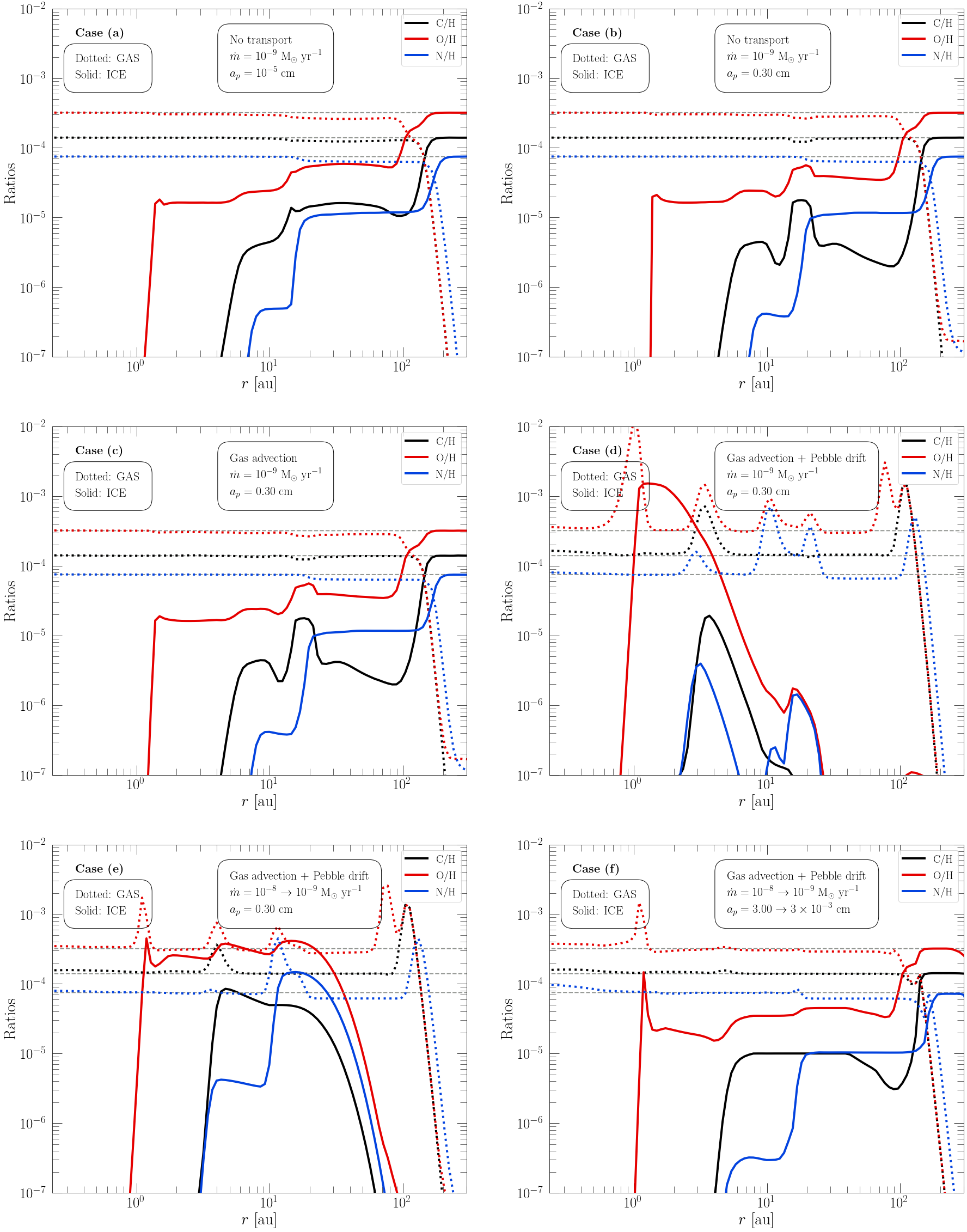}
    \caption{Final total C/H, O/H and N/H ratios after $10^5$ years. Dotted lines and solid lines correspond gas and ices, respectively. The gray dashed lines show the initial adopted Solar values for each of these ratios.}
    \label{fig:ratios}
\end{figure*}{}

\begin{figure*}
    \centering
    \includegraphics[width=1.0\textwidth]{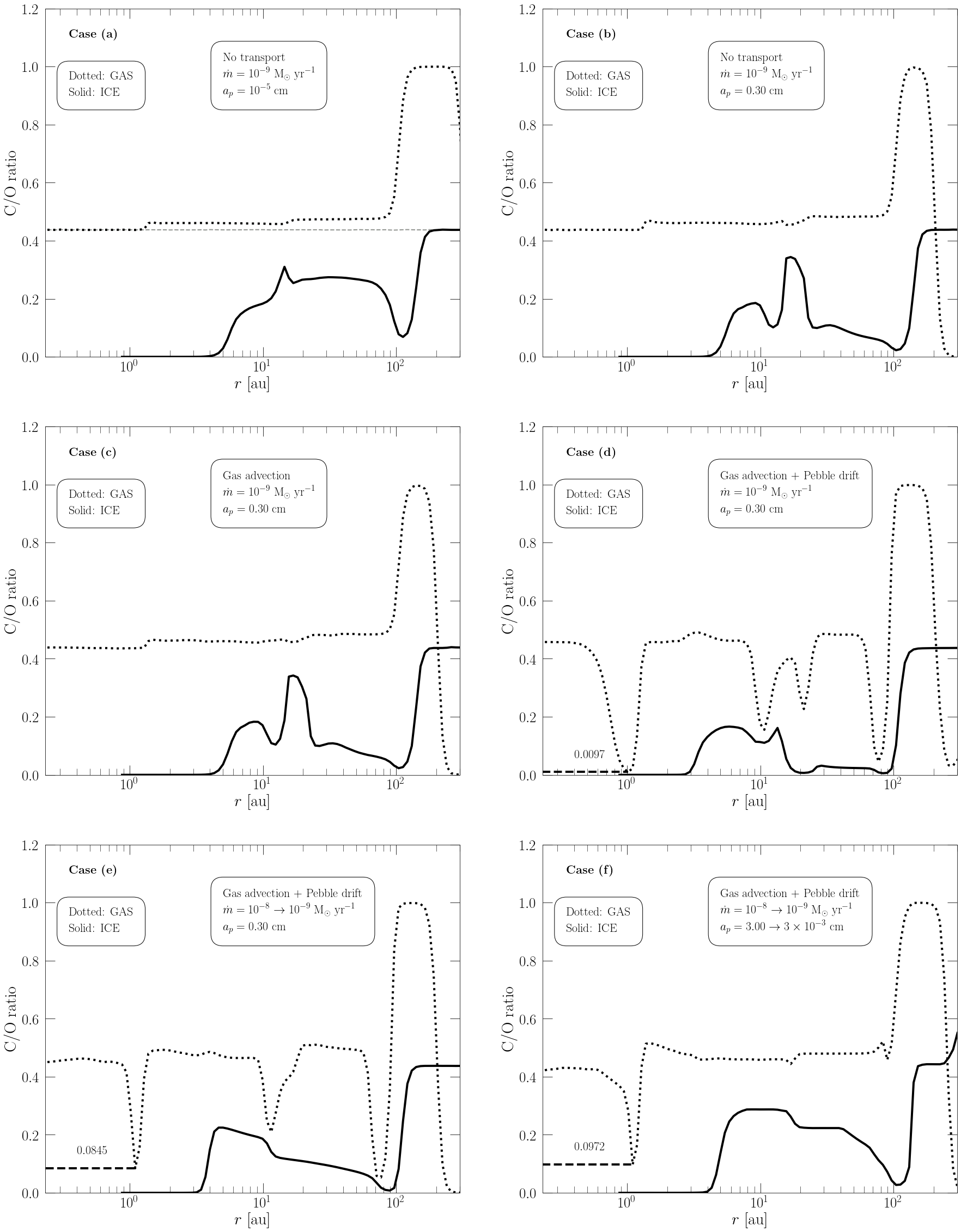}
    \caption{Final C/O ratio after $10^5$ years for Cases (a) to (f), as labelled. Dotted and solid lines correspond gas and ices, respectively. The gray horizontal dashed line shows the fiducial Solar value $\rm C/O_{\text{initial}} = 0.43$. The blue horizontal dashed line indicates the minimum value reached by the C/O ratio at the water iceline for models with radial drift of pebbles.}
    \label{fig:ratioCO}
\end{figure*}

\begin{figure}
    \centering
    \includegraphics[width=0.45\textwidth]{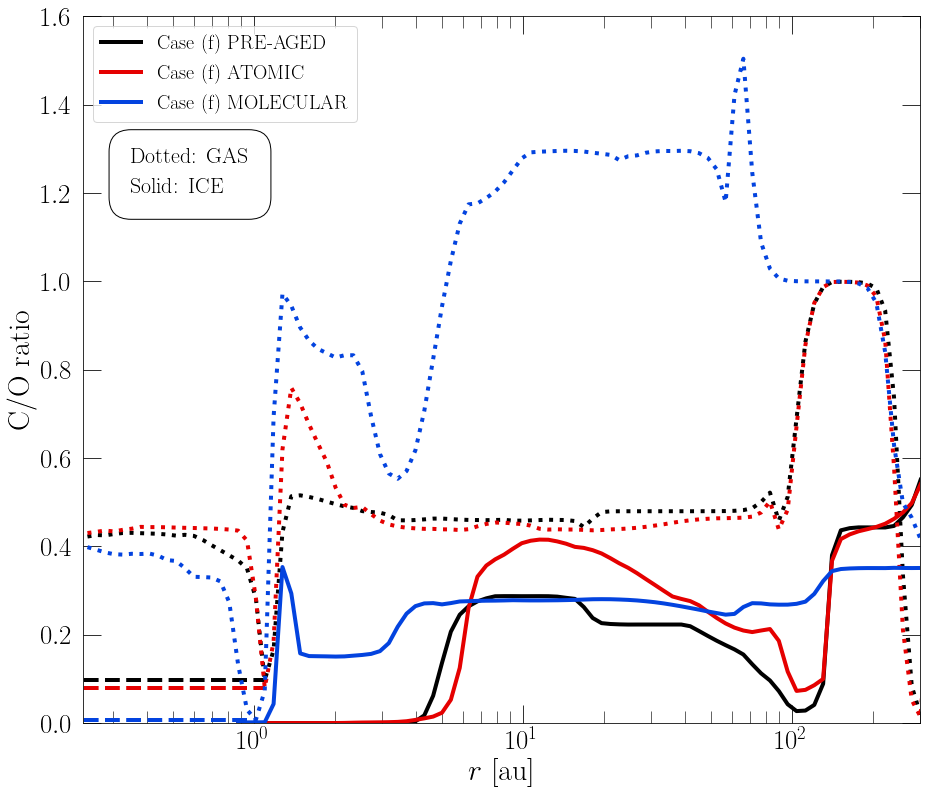}
    \caption{Comparison of the final C/O ratio after $10^5$ years for atomic and molecular variations of Case (f) with the fiducial pre-aged result. Dotted and solid lines correspond gas and ices, respectively. The horizontal dashed line indicates the minimum value reached by the C/O ratio at the water iceline. For the molcular case, this value is of the order of $10^{-3}$.}
    \label{fig:CO_extra}
\end{figure}

We have seen that the inclusion of advective mechanisms in a chemically evolving protoplanetary disk has significant consequences for the distribution of species in the gas and ice phases. This is also true for the total abundances of key elements, such as C, O and N. Figures \ref{fig:ratios}a---\ref{fig:ratios}f show the ratios of the total abundances for carbon, oxygen and nitrogen nuclei with respect to H nuclei in the gas (dotted) and ice (solid) for the six cases previously discussed. 
The thin horizontal dotted lines indicate the starting Solar abundance ratios. In cases \textbf{a}---\textbf{c}, where no radial drift is present, the total abundance of species at each radial point (both gas and ice phase) does not change; only the partitioning between the phases. 

By the radial evolution of both total element ratios and chemical species abundances results, eight regions can be characterized. These are:

\begin{enumerate}

    \item Beyond the $\rm N_2$ iceline ($\gtrsim 180\:$au), where there are no significant amounts of C, N, O present in gas phase, and ice-phase C, N, O abundances have fiducial Solar values.
    
    \item Between the $\rm N_2$ and CO icelines ($\sim 130\:$au). Here N achieves Solar abundance in the gas phase, while virtually all carbon and oxygen remain in the ice phase.
    
    \item Between the CO and $\rm O_2$ icelines ($\sim 100\:$au). C achieves near Solar abundance in the gas phase, while gas-phase O is present at a level $<50\%$ of Solar.
    
    \item Between the $\rm O_2$ and NO + HNO icelines ($\sim 15\:$au), where gas-phase O achieves near Solar abundance, with a modest fraction, $\sim 10\%$, left in the ice phase. At this point, ice-phase C/H and N/H reach comparable magnitudes.
    
    \item Between the shared NO + HNO icelines and the $\rm NH_3$ iceline ($\sim 7\:$au). Here the majority of C, N and O exist in the gas phase. In general, $\sim 1\%$ of the total N remains in the ice, compared to $\sim 4\%$ of C and $\sim 9\%$ of O.
    
    \item Small region between the $\rm NH_3$ iceline and $\rm CO_2$ iceline ($\sim 6\:$au). The effect of $\rm CO_2$ being in ice form has little effect on C/H and O/H ratios due to its largest abundance being almost two orders of magnitude lower than that of the also present CO and $\rm O_2$. This is a region where $\rm CO_2$ and $\rm H_2O$ are the main remaining species in the ice phase. As a consequence, ice becomes roughly one part carbon for three parts oxygen.
    
    \item Between the $\rm CO_2$ iceline and the $\rm H_2O$ + $\rm H_2O_2$ icelines ($\sim 1.5\:$au), where the gas reaches near solar abundances, except for modest deficit of O due to $\rm H_2O$ and $\rm H_2O_2$ ices. This results in an overwhelming presence of O, compared to C and N.
    
    \item Between the $\rm H_2O$ + $\rm H_2O_2$ icelines and the DZIB ($\sim 0.1\:$au to 0.7~au, depending on accretion rate). Here, even though chemistry here is very active, with a variety of abundances for many species, virtually all this takes place in the gas phase; therefore, all atomic ratios for gases are those of the adopted Solar abundances.
    
\end{enumerate}

When radial drift of pebbles is active in cases \textbf{d} to \textbf{f} the total element ratios reveal the enhancement of gas-phase volatiles at the icelines and the depletion of ices due to transport. The boundaries of the above regions can be more readily discerned, especially for Case \textbf{d}, which is the model with overall fastest transport of pebbles. Super-solar gas-phase abundances of O and C can be achieved, especially for O at the $\rm H_2O$ iceline.


\subsection{C/O ratios}

Figures \ref{fig:ratioCO}a to \ref{fig:ratioCO}f show the C/O ratios for gas (dotted) and ice (solid) for the same cases previously reported. All species in the chemical network are being considered, not only the main reservoirs. The thin horizontal dashed lines indicate the canonical Solar C/O ratio at $0.43$. The blue dashed line shows the minimum C/O ratio achieved at the water iceline. Looking at the fiducial case, the partitioning of C- and O-bearing species between gas and ice phases gives way to a ``step-like function'' for the gas-phase ratio. Aside from the small peak at $\sim 13$~au and a big dip at $\sim 100$~au that can be attributed to the unequal creation rate of ice-phase $\rm CO_2$ at different locations, the ice-phase C/O ratio also shares this step-like behavior. This resembles the C/O ratios previously described in \cite{Oberg_2011} (see their Figure 1), which involves the C/O ratio depending solely in the locations of the icelines. However, their model assumed $\rm H_2O$, $\rm CO_2$ and CO to be the main carriers of C and O atoms. Unless CR-ionization is suppressed we find $\rm O_2$ to be an important oxygen reservoir, regardless of initial conditions. Atomic and pre-aged scenarios also develop a significant abundance of NO + HNO (see Figures \ref{fig:abu1_NT} and \ref{fig:abu2_NT}. Similarly to the atomic ratios, detailed chemical models allow us to update the C/O ratio in a variety of protoplanetary disks as being controlled by the locations of the following icelines:
\begin{enumerate}
     \item Beyond the $\rm O_2$ + CO icelines ($\gtrsim 180$ au).
    \item Between the approximate location of the $\rm O_2$ + CO icelines ($\sim 130$---$180$~au) and the NO + HNO icelines ($\sim 10$---$20$~au).  
    \item Between the NO + HNO icelines ($\sim 10$---$20$ au) and the $\rm CO_2$ iceline ($\sim 6$~au) 
    \item Between the $\rm CO_2$ icelines ($\sim 6$ AU) and the $\rm H_2O$ + $\rm H_2O_2$ icelines ($\sim 1.5$~au)
    \item Between the $\rm H_2O$ + $\rm H_2O_2$ icelines ($\sim 1.5$~au) and the DZIB.
\end{enumerate}

When radial drift of pebbles is turned on (Cases (d) to (f)) the C/O ratio reflects the the enhancement of gas-phase volatiles at their icelines. The high abundance of oxygen-rich volatiles produces a profile with a number narrow ``valleys'' proportional to the existing enhancement. Only at the location of the $\rm CO_2$ iceline does the ratio shows a (very slight) enhancement.  Case (f), which is considered the most realistic, shows most of these enhancements to be minor, except at the water iceline. With significant pebble flux, the C/O ratio has dropped below 0.1. The limited evolution of the dust-to-gas mass ratio (see Figure \ref{fig:dustratio}) for this case and the extremely low value C/O$<0.01$ reached by faster Case (d) hints that this value will drop further if the system is evolved for a longer time. In addition, as time advances further, gas advection is expected to spread the O-rich gas inwards towards the DZIB region.

In Figure \ref{fig:CO_extra} we explore the C/O ratio evolution for two more variations of Case (f), in which we run models with atomic and molecular initial abundances as described in Table \ref{tab:initial}. The resulting abundances of these alternate cases can be found at the bottom rows of Figures \ref{fig:abu1_NT} and \ref{fig:abu2_NT} in Appendix \ref{sec:appendix}. Compared to the model with chemically pre-aged initial and outer injection abundances, the model with atomic abundances has only relatively modest repercussions on the overall gas-phase C/O ratio profile.
First, the ice-phase C/O ratio shows slightly increased values in regions between 100~au and 10~au, since water ice is less abundant overall, while $\rm CO_2$ ice remains reaches similar levels to the pre-aged model. 
Second, the gas phase abundance of $\rm O_2$ decreases more rapidly inside $\sim 3\:$au 
leading to a moderately higher gas phase C/O ratio in the atomic model at $\sim 2\:$au, i.e., just exterior to the water ice line.
%
For both pre-aged and atomic models, the ice phase C/O ratio drops towards values less than $10^{-5}$ interior to the $\rm CO_2$ iceline since only traces of carbon bearing ices remain, i.e., mainly methanol.

The molecular scenario shows greater changes in its C/O gas and ice profiles. $\rm H_2O$, $\rm CO_2$ and especially $\rm CH_4$ are introduced to the disk at very high abundance compared to the pre-aged model. After the CO iceline is crossed, sublimation of $\rm CH_4$ at 70~au allows the gas-phase C/O to reach its highest value of 1.5. Further interior in the disk, more abundant $\rm CH_4$ gas keeps the gas-phase C/O value high at around 1.3, before the sublimation of the much more abundant $\rm CO_2$, and later $\rm H_2O$, decreases the ratio in steps. Just interior to the $\rm CO_2$ iceline in particular, there is a localized enhancement of gas phase $\rm CO_2$ due to advected delivery of pebbles, which viscous spreading in the disk has not had time to spread out over the inner disk regions. This causes a local minimum in the gas phase C/O ratio, but then returning to values $\gtrsim 0.8$ inside about 3~au. Another major difference is apparent in the ice phase, where a $\sim 10^{-5}$ abundance of $\rm CH_3OH$ permits the ice to keep a C/O ratio of around 0.2. In the pre-aged and atomic models, this ratio in the ice drops down to $< 10^{-4}$. The iceline of $\rm CH_3OH$ occurs just before water and provokes both the C/O ratio increase in the gas and ice seen at about 1.2~au, i.e., due to enhanced absolute abundances of both gas and ice phase $\rm CH_3OH$ due to advection. Finally, just interior to the water ice line, the gas phase C/O ratio now decreases to much smaller values, i.e., $\sim 10^{-3}$, than the pre-aged and atomic models, which reach $\sim 0.1$.


\section{Discussion} \label{sec:dis}

\subsection{Impact of transport mechanisms for disk composition}

We have investigated how the abundances of the main carbon, oxygen and nitrogen-bearing species in the midplane of a protoplanetary disk evolve under the effects of a detailed chemical network, radial transport, evolving accretion rate, and changing pebble size, with the evolution followed for $10^5$ years.

Gas advection continuously shifts gas-phase species to inner regions, reducing the time a gas parcel would spend in place and be subjected to local chemical processing. However, an almost flat velocity profile and a vast distance to cover ensures chemical evolution dominates in the outer regions. When entering the active region ($\sim 2$~au for a $\dot{m} = 10^{-9}\:M_{\odot}\:{\rm yr}^{-1}$ disk), gas speed picks up and, thanks to enhancements provided by radial drift, can more effectively compete against chemical kinetics. 

Radial drift of pebbles acts by quickly transporting molecular species frozen onto the grains to inward regions. These then sublimate at their icelines and greatly enhance the gas-phase abundance. If the icelines are located in the active region, as is the case for $\rm H_2O$ and $\rm CO_2$, gas advection has an enhanced ability to spread the oxygen-rich gas (and associated low C/O ratio) to reach the DZIB before dissipation of the disk after 1 to 2 Myr. Another consequence of pebble drift is that the dust-to-gas mass ratio drops considerably in the outer region. 

Interestingly, \cite{Banzatti2020} have studied 63 protoplanetary disk by spatially resolving the distribution of dust/pebbles from millimeter continuum ALMA observations and examining IR emission that traces the warm inner disk atmosphere as observed by \textit{Spitzer}. They found an anti-correlation between the $\rm H_2O$/HCN ratio and the dust disk radius $R_{\text{dust}}$ in a number of protoplanetary disks of ages 1---3 Myrs and various sizes. This suggests the existence of efficient pebble drift that depletes the outer disk of dust and enriches the inner regions of these disks with water vapor. Those disks maintaining a large $R_{\text{dust}}$ are thought to be indicative of the presence of dust traps that prevent migration of the pebbles. Large disks with inner cavities, where little molecular gas is present, could be explained by inefficient drift in the outer region, but with efficient drift in regions closer to the star. 

The resulting composition of the disk is affected by its previous history, i.e., evolving from a state with a greater accretion rate. Apart from the slower depletion of the dust component, the changing temperature of the active region induces a horizontal spatial spread of the icelines. This provokes a more restrained enhancement of molecules (e.g., $\rm H_2O$) due to slower drift velocities at higher accretion and by it being spread out at various points in the disk, rather than a very large enhancement confined to just one region \citep[see, e.g.,][]{gavino2021impact}.

While transport mechanisms have strong influence, chemical processing can still be the controlling process. 
The midplane composition is sensitive to the choice of CRIR; even more so than initial conditions \citep[see, e.g.,][]{cleeves2013,Booth_2019}. In addition, the introduction of the pebble radial size profile considerably weakens the effects of radial drift for regions with small pebble sizes (found at $>10$~au in our case). Slow transport heightens the effect of chemical processing, specially in cases with high CRIR \citep[see][]{Booth_2019}.

Making a more specific comparison to the results of \cite{Booth_2019}, we see that their 1-Myr, $\alpha = 10^{-3}$, pebble drift models show icelines with a more moderate, i.e., one order of magnitude, final enhancement for $\rm CO_2$ abundance and no noticeable enhancement for $\rm H_2O$, with respect to their initial abundances. We note that in these models gas advection has had more time to spread enhanced gas phase abundances inward from the ice lines. Their higher initial abundances of $\rm CO_2$ and $\rm CH_4$, at the expense of CO, allows for significant carbon-rich gas to spread inwards and mix with the remaining water rich gas in the inner region. As a result, at 1 Myr their gas-phase C/O ratio has climbed back up from close to zero at $50 \times 10^3$ years to approximately 0.8. This comparison highlights how results of such disk modeling can vary due to differences in assumptions for both physical and chemical aspects of the models.


\subsection{CO abundances and C/O ratios}

In general, all models presented here with pre-aged initial/boundary conditions show a relatively stable CO gas abundance of $\sim 10^{-4}$, which changes little over the modeled time interval. Evolution to longer timescales ($\sim 1\:$Myr) leads to some CO gas depletion, but only by factors of a few and in specific locations. The presence of large enough grains/pebbles and associated strong radial drift creates enhanced CO gas abundance just interior to the CO ice line, i.e., in Cases (d) and (e) shown in Figure~\ref{fig:abu}. However, this feature is not significant in Case (f) due to its smaller grains/pebbles in the outer disk.
As for C/O ratios, models with pre-aged initial conditions always keep a gas phase C/O~$<1$ inside the CO iceline, while models computed using molecular initial conditions are able to reach gas phase C/O $>1$ between 10 and 100 au. However, these will drop in the same way as the pre-aged and atomic scenarios as one reaches the $\rm CO_2$ and water icelines. Therefore, all explored cases end up having gas phase C/O~$<1$ at radii $<10$~au. Such gas phase CO abundances and C/O ratios results are similar to those reported by \cite{Booth_2019}.

These results conflict with ALMA observations of $\rm C_2H$ column densities for some protoplanetary disks, which imply elevated C/O ratios (i.e., $\sim 2$) and depleted gas phase CO abundances (by up to a factor of 100) within the CO iceline \citep[see, e.g.,][]{Bosman_2021}. We consider that reconciliation of these observational results with our theoretical models may involve some combination of the particular regions of the disk that are being probed by the observations and the evolutionary stage of the disks.

Regarding vertical disk structure, it has been shown that the upper layers of a disk can become depleted of CO gas via vertical settling of CO ice coated grains towards the midplane. Such settling, including water ice, can also increase the C/O ratios of these upper layers. At the same time, the midplane becomes the main reservoir of CO-ice, resulting in a different composition to that of the upper layers \citep[see, e.g.,][]{Krijt_2018}. Observations of gas phase species can be biased to probing conditions of the warmer surface layers, so care must be taken in assuming their results hold for the global disk properties averaged over the vertical structure. Furthermore, when connecting to planet formation it is expected that planets accrete relatively little mass from the disk surface layers, with planetary C/O ratios being mostly set by disk midplane conditions \citep[e.g.,][]{Cridland_2020,johansen2021pebble}.

Regarding evolutionary stage, there is some evidence of disks that have relatively low C/O ratios, e.g., C/O~$\sim$ 0.8 in IM Lup \citep[][]{Ilsedore_Cleeves_2018}, which is thought to be at a relatively early evolutionary stage. Low C/O ratios may thus be a feature of younger disks that are in the process of transforming their volatile and O-rich gas to have volatile poor and C-rich compositions.

\subsection{Implications for planetary compositions}

\subsubsection{Planetary cores}

Inside-Out Planet Formation (IOPF) theory \citep{CT14} proposes that planets form from pebbles that drift from the outer disk and that are trapped at the pressure maximum associated with the dead zone inner boundary (DZIB). The location of the DZIB is set by thermal ionization of alkali metals, i.e., Na and K, that occurs at temperatures around $1,200\:$K \citep[see also][]{Jankovic_2021}. In the accretion-heated active region of the disk midplane, this temperature is reached at $\sim0.1\:$au when the accretion rate is $\sim 10^{-9}\:M_\odot\:{\rm yr}^{-1}$ and at $\sim 1\:$au when it is $\sim 10^{-8}\:M_\odot\:{\rm yr}^{-1}$.

\begin{figure*}
    \centering
    \includegraphics[width=1.0\textwidth]{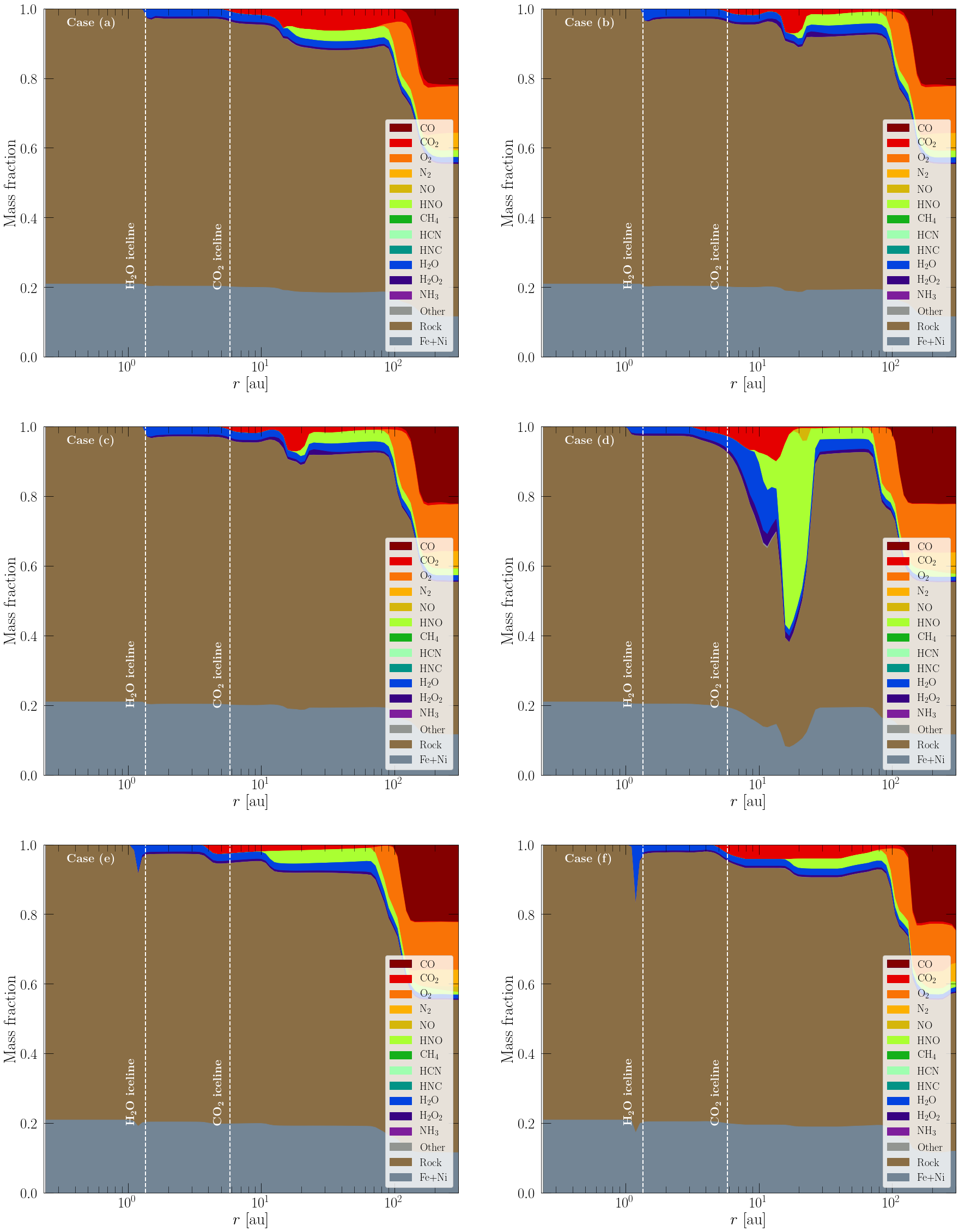}
    \caption{Mass fraction of each of the solid components in the disk: ice-species and grains. ``Fe+Ni'' stands for the fraction of the pebbles composed of bare iron and nickel (see text).}
    \label{fig:icerock}
\end{figure*}

A basic prediction of IOPF is that the planetary cores of the close-in Super-Earths / Mini-Neptunes should be very volatile poor. Such a prediction appears to be consistent with the inferred properties of these planets \citep[e.g.,][]{2021MNRAS.503.1526R}. Once inside the water iceline at about 1~au, pebbles are expected to have very little surface ice. This is displayed in Figure \ref{fig:icerock}, which shows the mass fraction of the ices and refractory components of the pebbles at each disk position after $10^5$ years.

\begin{figure*}
    \centering
    \includegraphics[width=1.0\textwidth]{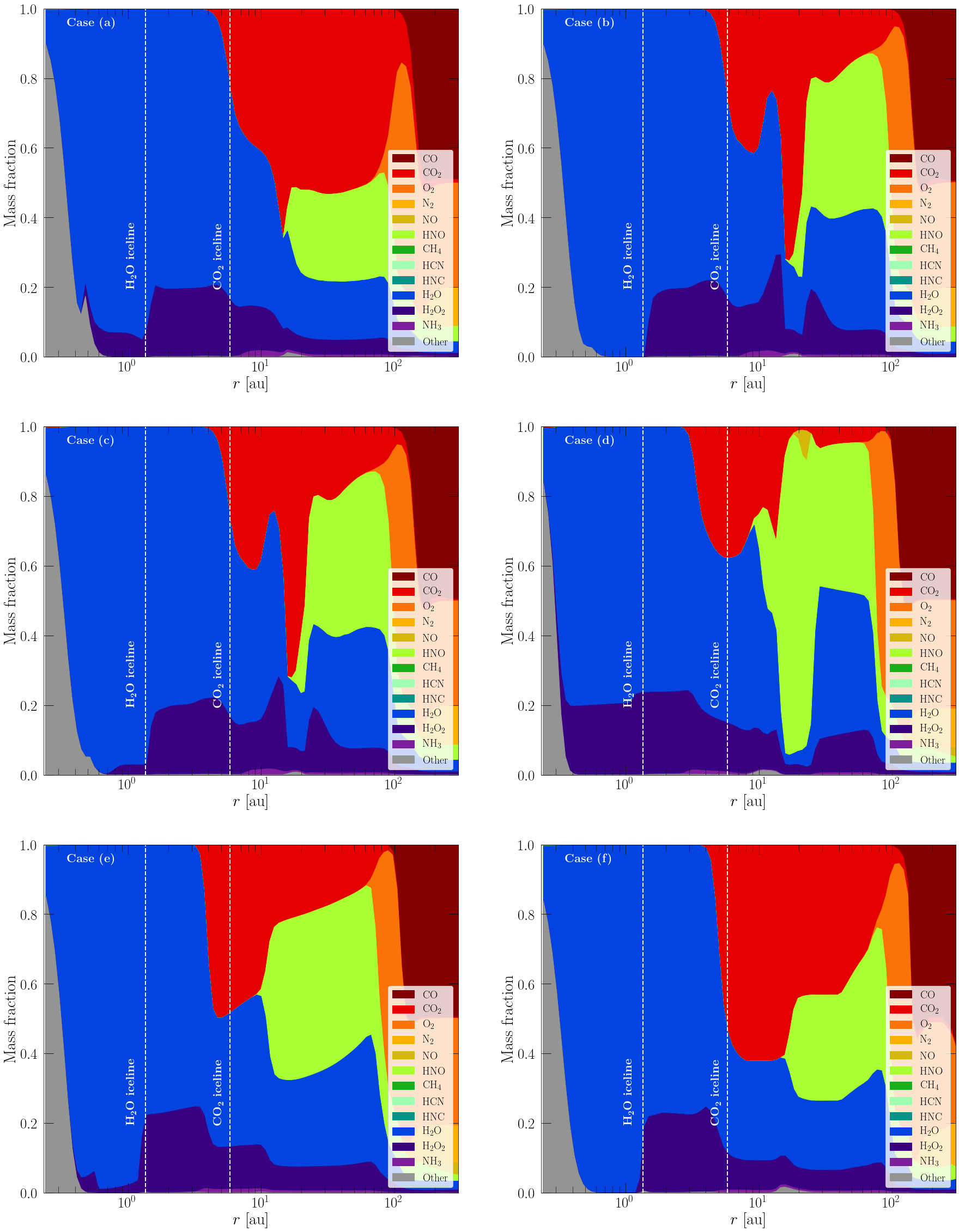}
    \caption{Mass fraction of each of the ice phase components in the disk.}
    \label{fig:ice}
\end{figure*}

While the innermost planets are expected to be volatile poor, IOPF also predicts that a chain of planets forms as the DZIB retreats outwards. At these later stages the DZIB location is potentially set by penetration of X-rays from the protostar \citep{Hu_2015} and the temperature at its location will be cooler than its initial value of 1,200~K. If the DZIB retreats to the vicinity of the water iceline and beyond, then the pebbles will be relatively water rich. The results of Case (d) implies that the mass fraction of water in the pebbles would be $\lesssim 3\%$. However, in Cases (e) and (f) we find localized enhancement of the ice mass fraction up to $\sim 15\%$ due to the advected delivery of pebbles and enhanced absolute abundances of $\rm H_2O$.
Note, in comparison with the estimated water mass fraction of Earth of $0.1\%$ \citep[see e.g.][]{Dishoeck2014}, all these levels of water content would likely be sufficient to create a circumplanetary ocean or ``water world''.
Beyond about 6~au, planets may also start receiving a significant fraction of $\rm CO_2$ ice (see Figure \ref{fig:ice}, which shows the mass distribution in the ice-phase component only). However, this is only at a level such that the total volatile contribution to mass is $\lesssim 5\%$.
We note that such small mass fractions would be difficult to infer from planetary bulk densities. Instead, as discussed below, the water or $\rm CO_2$ content of the atmospheres of the planets would be expected to be enhanced and more readily discernible, e.g., via transit spectroscopy.

From the Solar abundance estimates for carbon, oxygen and various minerals \citep[see][]{draine2011} and from our initial abundances given in Table \ref{tab:initial}, we estimate the Fe-Ni core mass fraction to be in the vicinity of $\sim 20\%$ in the inner regions. This value is close to the estimates for that of Venus and Earth \citep[see, e.g.,][]{Zeng_2016}. This mass fraction becomes lower if ices are present, with the lowest values for planets forming in the outermost regions of the disk. Increasing the prominence of the Fe-Ni core (e.g., as is seen in Mercury) would only happen in regions where the temperature would be high enough to vaporize the silicate component. Such conditions could exist in the very innermost regions of the disk, close to the initial DZIB that is at 1,200~K and thus relevant for the first, so-called ``Vulcan'', planet to form in the IOPF sequence. A study of these processes, including the modeling of the various ``rocklines'' \citep[e.g.,][]{Aguichine_2020}, is deferred to a future work.

\subsubsection{Atmospheres}

The chemical composition of the gas phase of the inner protoplanetary disk is also likely to be important for planets forming via IOPF. IOPF predicts that planets form with characteristic masses of $\sim 4 M_\oplus$, set by the process of shallow gap opening leading to truncation of pebble accretion \citep{CT14,Hu_2018}. Such planetary mass cores are expected to be able to accrete modest amounts of gas, i.e., $\sim 10^{-2}$ to $10^{-1}\:M_\oplus$ \citep[e.g.,][]{2014ApJ...797...95L}. Such gas accretion would not affect the mass of the planets very significantly, but would influence their size and thus lead to a range of bulk densities, such as is observed in the close-in Super-Earth / Mini-Neptune population \citep[e.g.,][]{2015ApJ...798L..32C}. For those planets retaining most of their primordial atmosphere, i.e., those resistant to photoevaporative losses, the chemical composition of their atmosphere is expected to retain memory of the initial composition inherited from the gas disk.

We see from our radial drift models, i.e., see Fig.~\ref{fig:ratioCO} Cases (d), (e) and (f), that accreted atmospheric gas from inside the water iceline could achieve C/O ratios that are very low, i.e., $\lesssim0.1$. Note that this minimum value is achieved at the water iceline, but is then expected to spread inwards due to gas advection. While the amount of such spreading is quite limited within the $10^5\:$yr evolution presented in Fig.~\ref{fig:ratioCO}, we note that this is quite sensitive to the adopted value of the viscosity parameter, i.e., $\alpha=10^{-4}$, and ignores any prior evolution from the earlier, higher accretion rate phases, e.g., in Case (f) the period when $\dot{m}>10^{-8}\:M_\odot\:{\rm yr}^{-1}$. Our limited treatment of only the disk midplane and neglect of vertical mixing also likely means that we have underestimated the rate of inward spreading. Furthermore, much of the gas accreted by the protoplanets may occur at later times, after their cores have formed, i.e., on timescales $\sim 10^6\:$yr when gas is still present in the disk. For all these reasons, we expect that the C/O ratio of much of the gas accreted by protoplanets forming {\it in situ} in the inner regions via IOPF will be relatively low, i.e., $\lesssim 0.1$.



\subsection{Caveats}\label{subsec:cav}

We have presented an extensive, but still simplified picture of the physical and chemical evolution of a protoplanetary disk. 
For the physical conditions, we have focused on midplane conditions thought to be relevant for most of the pebbles that grow large enough to undergo significant vertical settling. The model has also been presented only for a disk around a solar mass star and for a range of accretion rates near $\sim 10^{-9}\:M_\odot\:{\rm yr}^{-1}$, which are thought to be relevant to the IOPF model. STIP-systems are also routinely found around much lower-mass and cooler M dwarf stars \citep[see, e.g.,][]{Muirhead2015}. A future work will study the physical and chemical properties of disks around such stars.


We also note that we have adopted only one cosmic ray ionization rate (CRIR) in our main model series, i.e., $\zeta=10^{-17}\:{\rm s}^{-1}$. Higher levels of cosmic-rays induce a higher rate of chemical processing (see Appendix \ref{sec:appendix}). The CRIR is very uncertain and indeed may vary from system to system, including due to location with the Galaxy. Contributions to the ionization rate from locally produced X-rays, potentially relevant for the inner disk regions, have also not been included. As shown in Fig.~\ref{fig:abu1_NT}, one of the main effects of an enhanced CRIR is to increase the abundance of $\rm CO_2$ ice in the outer disk at the expense of gas-phase CO. However, the results for the inner disk regions and thus for IOPF are relatively insensitive to this difference.


Another caveat concerns our fiducial choice of timescale over which to consider the evolution, i.e., $10^5\:$yr. This is an order of magnitude less than those reported by similar works \citep[e.g.,][]{Eistrup_2016,Booth_2017, Booth_2019}, being well within the lifetime of nearby young protoplanetary disks \citep[see, e.g.,][]{williams2011}. However, our choice is motivated by the desire to follow the chemistry to the point of first planet formation within the IOPF scenario and we have also included a period of chemical pre-aging of the initial conditions. Furthermore, given the high densities of the disk, our basic chemical results are relatively insensitive to the choice of this timescale. 

For models with radial drift, longer evolutionary timescales of 1 Myr, which are presented in Appendix \ref{sec:appendix_1Myr}, show enhanced abundances of certain gas phase species due to enhanced delivery of icy pebbles to various icelines. As also discussed here, the choice of outer disk radius is also quite arbitrary. While we have presented models that extend out to 300~au, such large disks can become unrealistic in terms of their total mass, especially for the high accretion rate case with $10^{-8} M_\odot\:{\rm{yr}}^{-1}$. There is scope to improve the ``toy'' models presented here to include more realistic global evolution that could involve different prescriptions for the time evolution of the outer disk radius, radial variation in the magnitude of disk viscosity, and a greater exploration of different accretion histories. Such explorations are also deferred to a future study.

The pre-aging of initial abundances serves to represent chemical processing taking place as dissociated gas atoms and grains (reset-scenario) move from the molecular cloud phase to the disk midplane. However, the timeframe, temperatures and densities of these regions, and ionization rates are quite uncertain. The timeframe of $10^4$ years was chosen as to allow the most important ice-phase species CO, $\rm CO_2$ and $\rm H_2O$ to reach quasi-stable abundances. Allowing for longer evolution has little effect on our results. We have also considered the effects of changing the initial and injected outer boundary abundances to example ``Atomic'' and ``Molecular'' models (see Appendix~\ref{sec:appendix}). The atomic model shows only modest differences compared to our fiducial pre-aged model, especially in the C/O ratios. However, the molecular model has some significant differences, especially related to more abundant $\rm CH_3OH$ ice that keeps ice phase C/O relatively high at $\sim 0.2$ inside 6~au and spiking close to 0.4 just exterior to the water ice line. Also, in this model the gas phase C/O is driven down to a much lower minimum value $\sim 10^{-3}$ at the water ice line, which is then expected to spread inwards, and so may be discernible from observations of atmospheres of close-in planets.

Concerning the physical dynamics of gas and grains, by having a midplane model we ignore phenomena that could be explored in a two-dimensions, such as vertical mixing, volatile enhancements in icelines that have vertical structure, and dust settling. All of these effects may have some influence on the chemical history of any potential planet. Future work to extend the modeling to two and three dimensions is warranted.

Finally, grains/pebbles are treated as having a single representative size for each disk location. In reality there would be a distribution of sizes \citep[see, e.g.,][]{Hu_2018}. However, our final model presented in Case (f) does allow for size variation with radius and its radial profile has been guided by the results of pebble distributions modelled by \cite{Hu_2018}. Nevertheless, pebbles of different sizes would undergo radial drift at different speeds and this could induce systematic differences in the resulting chemical abundances. Again, accounting for this effect, which would entail much greater computational cost, is deferred to a future study.

\section{Conclusions}\label{sec:con}

The models presented in this work have demonstrated the importance of detailed chemical evolution coupled with physical transport phenomena on the chemical composition in protoplanetary disks midplanes. A particular focus was placed on the inner regions $\lesssim 1$~au, which is a relevant region for Inside-Out Planet Formation (IOPF). Our conclusions are:

\begin{enumerate}
\item Pebble drift enriches the inner disk with solids, while depleting the outer disk. A quasi-equilibrium D/G ratio can be reached in the outer disk regions, depending on the rate at which new material arrives from the outer boundary and the timescale over which the evolution is considered. However, the inner regions are unlikely to have reached an equilibrium, so that our presented results for abundances have some sensitivity to the choice of initial conditions and timescale of disk evolution.

\item Partitioning between ice and gas species is determined mostly by iceline positions. However, pebble drift transports ice from the outer regions to higher temperature zones. This enhances the gas phase abundances of volatile species at their icelines. The strength of the enhancement depends on the drift time $t_{\rm drift}$ of the pebbles, which generally favors cases with larger pebbles and lower gas densities.

\item Gas-grain reactions of gas-phase CO continuously supplies the protoplanetary disk with $\rm H_2O$ ice, which then drift towards its iceline where it sublimates and enhances water vapor abundance by more than two orders of magnitude. Slower but still significant gas transport is then expected to distribute the $\rm H_2O$ enhanced gas throughout the inner region $< 1$~au on timescales greater than $10^{5}$ years.

\item Accretion rate evolution introduces density and temperature changes, which shift the ice locations in the active region and modify pebble velocities. The consequences of this are reflected on the resulting D/G ratio and molecular composition of the midplane after $10^{5}$ years.

\item Introducing a pebble size profile, where the average size of grains increases the closer they are to the DZIB, modifies the D/G ratio behavior. Our chosen profile resulted in a very mild D/G increase at $\sim 2$ AU followed by a decreasing D/G in the innermost regions, where the grains are the largest. At the same time, at locations 10~au and greater the D/G remained virtually unchanged. This in turn reduces the importance of most volatile iceline enhancements, except that of $\rm H_2O$.

\item Due to being dominated by water, the C/O ratio of ices in regions $<5$~au is extremely low at $\sim 0.01$.
In all models featuring pebble drift, the gas-phase C/O ratio at the water iceline reaches values of $\lesssim 0.1$. The aforementioned expected spreading of this $\rm H_2O$ enrichment to innermost regions dust to continued gas advection implies that accreted planetary atmospheres in the inner disk, relevant for IOPF, should also have this low value of C/O.


\item Planets forming near the DZIB inside about 1~au would be volatile-poor, including water-poor, but could accrete a primordial water-rich atmosphere. With DZIB retreat, those forming at locations closer to the water iceline may take up a non-negligible amount of water ice, up to $\sim10$ percent of the planet's total mass.

\end{enumerate}

Future work is needed to explore further the effects of various simplifications and caveats associated with the presented models, including effects of chemical model uncertainties, ionization rates, pebble size distributions, choices of initial and boundary conditions, and approximation of a vertically thin geometry. Nevertheless, this work represents the first predictions for chemical compositions of planets forming in the Inside-Out Planet Formation scenario and the basic results may be compared and contrasted with those of other planet formation scenarios, such as outer disk formation followed by inwards tidal migration \citep[][]{2017MNRAS.469.4102M}.


\section*{Acknowledgements}
We thank an anonymous referee for comments that helped to improve the paper. We also thank Christian Eistrup for helpful discussions. This work is based on that submitted in partial requirements for Mr. Arturo Cevallos Soto's Masters thesis in Physics at Chalmers University of Technology, supervised by Prof. Jonathan C. Tan. We acknowledge Ms. Ana Mari Petrova for serving as the student opponent for the Masters defense.

\section*{Data Availability}
Data from the computed models is available upon request to the authors.



\bibliographystyle{mnras}
\bibliography{bibliography} 




\appendix

\section{Extended evolution model}\label{sec:appendix_1Myr}

\begin{figure}
    \centering
    \includegraphics[width=0.45\textwidth]{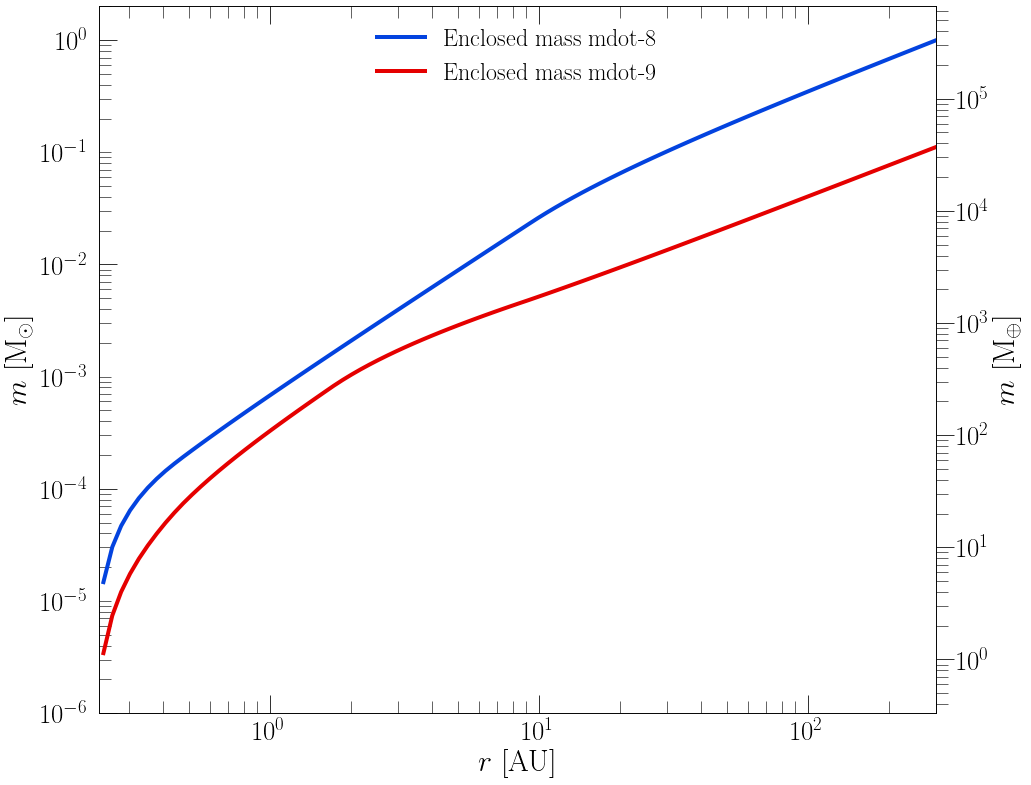}
    \caption{Enclosed gas mass profiles for the models with $10^{-8}$ (blue line) and $10^{-9}\:M_\odot\:{\rm yr}^{-1}$ (red line) accretion rates.
    }
    \label{fig:contained_mass}
\end{figure}

Here we present a model of disk physical and chemical evolution that has a ten times longer duration, i.e., $10^6\:$yr, compared to the fiducial case presented above that is evolved for $10^5\:$yr. As discussed in \S\ref{sec:dis}, the appropriate timescale for the evolution to the point of first planet formation in the IOPF model is quite uncertain, also noting that the starting condition we have chosen in Case (f) is already advanced to the point when the disk has an accretion rate of $10^{-8}\:M_\odot\:{\rm yr}^{-1}$ and that chemical pre-ageing of $10^4\:$yr has been applied to the abundances supplied at the outer boundary. Nevertheless, as discussed below, certain global properties of the disk, especially if it does extend to large radial size, may motivate an evolutionary timescale that is longer than $10^5\:$yr.

Figure \ref{fig:contained_mass} shows the enclosed disk mass, $m_{\rm disk}(<r)$, as a function of $r$ for models with accretion rates of $10^{-8}$ and $10^{-9}\:M_\odot\:{\rm yr}^{-1}$. As discussed in \S\ref{subsec:cav}, if the disk extends to 300~au and maintains a low value of $\alpha=10^{-4}$, then the total implied mass in the disk can be unrealistically large, i.e., about 0.1 to 1~$M_\odot$. For example, the latter value, being similar to the central stellar mass, would be expected to be gravitationally unstable. A more realistic model with smaller disk mass could involve the disk being smaller and/or having a more compact dead zone region, i.e., so that the outer disk has higher viscosity and thus smaller mass surface density. We consider that investigating these various possibilities is beyond the scope of the current paper, where we have focused on presenting a first exploration of the chemical evolution of simple, constant-$\alpha$ disks that are fiducial models expected to be applicable at least in the inner disk regions relevant to IOPF.

Next consider the time evolution of the decaying accretion rate disks, i.e., Case (f). Our fiducial model involves this accretion rate declining from $10^{-8}$ to $10^{-9}\:M_\odot\:{\rm yr}^{-1}$ over a period of $10^5\:$yr. However, this would not be consistent with the total mass of the disk enclosed within 300 au being transferred to the star at these accretion rates. Self-consistent solutions to this aspect could involve: (a) the disk being much smaller than 300 au; (b) significant mass being lost from the disk via a wind; (c) significant mass being accreted in disk surface layers, rather than in the midplane; (d) a more extended period over which the accretion rate declines.

\begin{figure}
    \centering
    \includegraphics[width=0.45\textwidth]{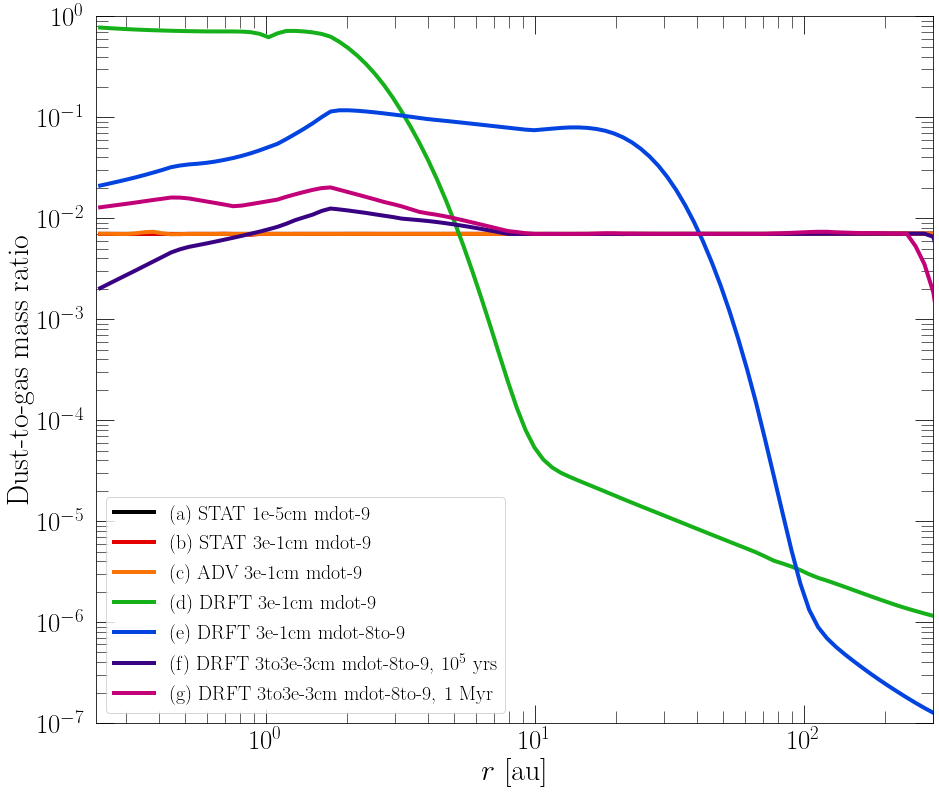}
    \caption{Final dust-to-gas ratio of Case (g) disk model, including comparison to other cases, as labelled. 
    }
    \label{fig:DG_compare_f}
\end{figure}

Here we proceed to investigate the last of these possibilities by increasing the duration of disk evolution to $10^6\:$yr.
Longer evolutionary timescales have consequences for both the radial distribution of pebbles in the disk and the disk's chemical composition. We present a new Case (g), which keeps the same initial conditions as Case (f), but has its accretion rate decay from $\dot{m}=10^{-8}\:M_{\odot}\:{\rm yr}^{-1}$ to $10^{-9}\:M_{\odot}\:{\rm yr}^{-1}$ over a period of 1~Myr.
Figure \ref{fig:DG_compare_f} compares the dust-to-gas mass ratio of Case (g) with those of the other cases. In particular, the longer evolutionary timescale of Case (g) causes the dust-to-gas ratio to become moderately larger than that of Case (f) inside about 10~au. This is the region of the disk where the pebbles are large enough that radial drift is occurring at significant speeds.

\begin{figure*}
    \centering
    \includegraphics[width=1.0\textwidth]{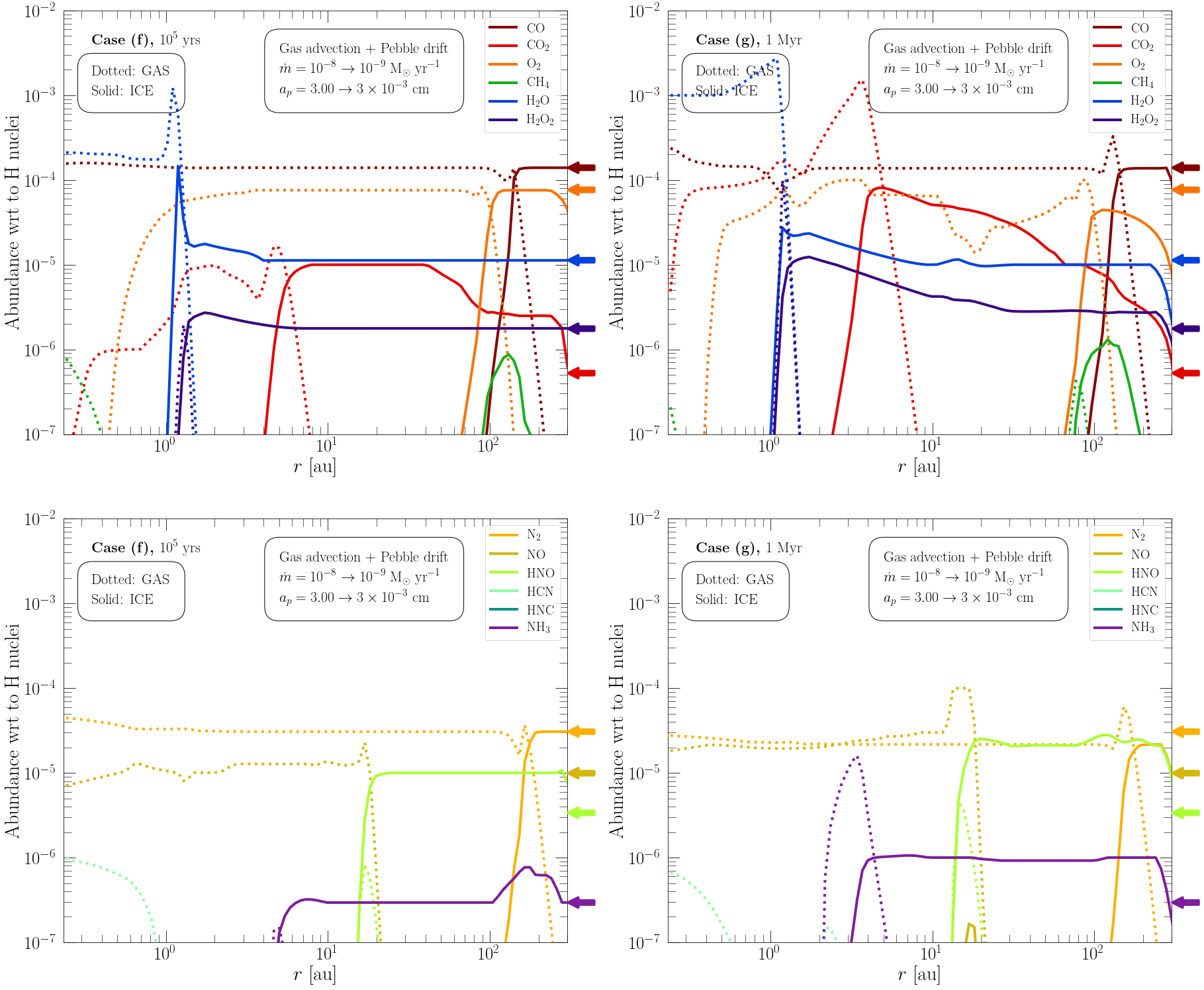}
    \caption{\textbf{Left column:} Final abundances for Case (f) (i.e., evolved for $10^5$ years; same as shown in Figs.~\ref{fig:abu} and \ref{fig:nitro}). \textbf{Right column:}: Final abundances for Case (g) (i.e., evolved for 1 Myr).}
    \label{fig:abu_compare_f}
\end{figure*}

Figure \ref{fig:abu_compare_f} compares the chemical abundances of Case (f) and Case (g).
The abundances of gas phase $\rm H_2O$, $\rm CO_2$ and $\rm NH_3$ show particular enhancements (sometimes by more than an order of magnitude) in the regions inside their icelines. This is partly due to there being more time for gas advection to spread inwards from the iceline region and partly due to some enhancements of the ice phase species delivered to these regions.

\begin{figure}
    \centering
    \includegraphics[width=0.45\textwidth]{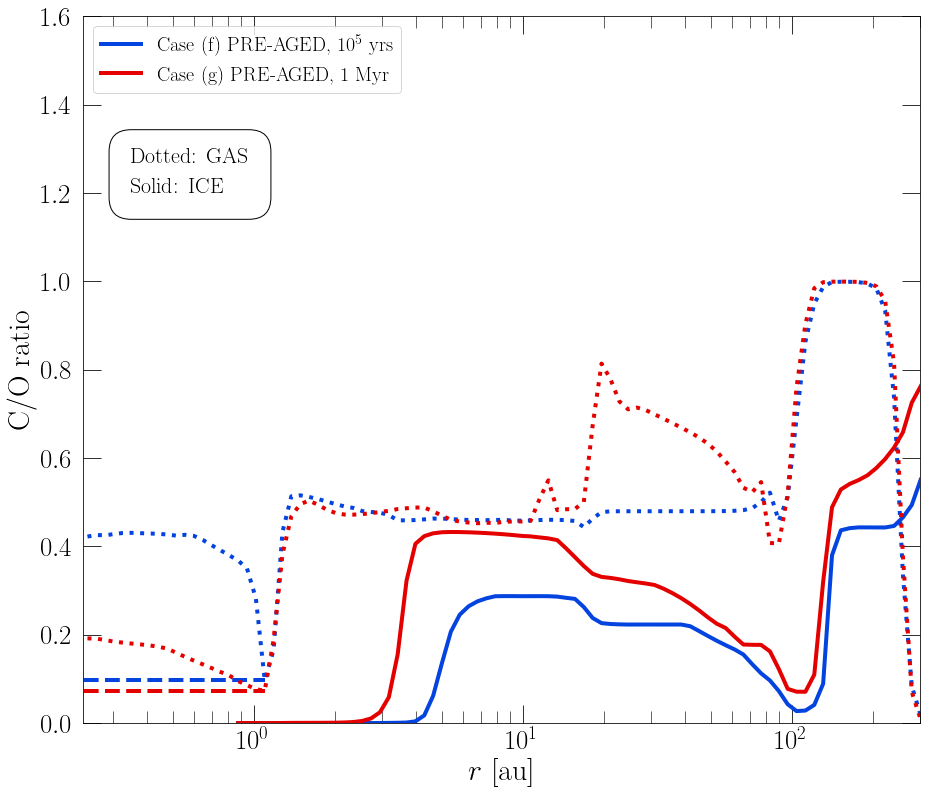}
    \caption{Final C/O ratios (gas phase - dotted lines; ice phase - solid lines) of Case (f) (blue) and Case (g) (red).}
    \label{fig:CO_compare_f}
\end{figure}

These compositional changes affect the C/O ratio (see Figure \ref{fig:CO_compare_f}). 
At the water iceline, the gas-phase C/O ratio in Case (g) at 1 Myr remains close to $\sim 0.1$, similar to the value here in Case (f) at 0.1~Myr. However, now with the spread of water vapor inwards via gas advection, the C/O ratio is lowered to $\lesssim 0.2$ throughout most of the region inside 1~au.
Further out, from 15 to 70 au, the gas-phase C/O ratio of Case (g) is larger (up to a value of about 0.8 near 20~au) than that of Case (f), which is due reduced abundances of $\rm O_2$ in the older disk. 
Concerning the ice-phase C/O ratio, the greater abundance of $\rm CO_2$ ice leads to enhancements of this ratio in Case (g) compared to Case (f) over a broad region from about 3~au to 100~au. Further out there is a similar enhancement in ice-phase C/O ratio of Case (g), but this is caused by its reduced $\rm O_2$ ice abundance.
This suggests that continued evolution would result in further enhancement of the ice phase C/O ratio due to these effects.

{}

{}

{}

\section{Effect of initial/boundary conditions and CRIR on disk composition}\label{sec:appendix}

\begin{figure*}
    \centering
    \includegraphics[width=0.95\textwidth]{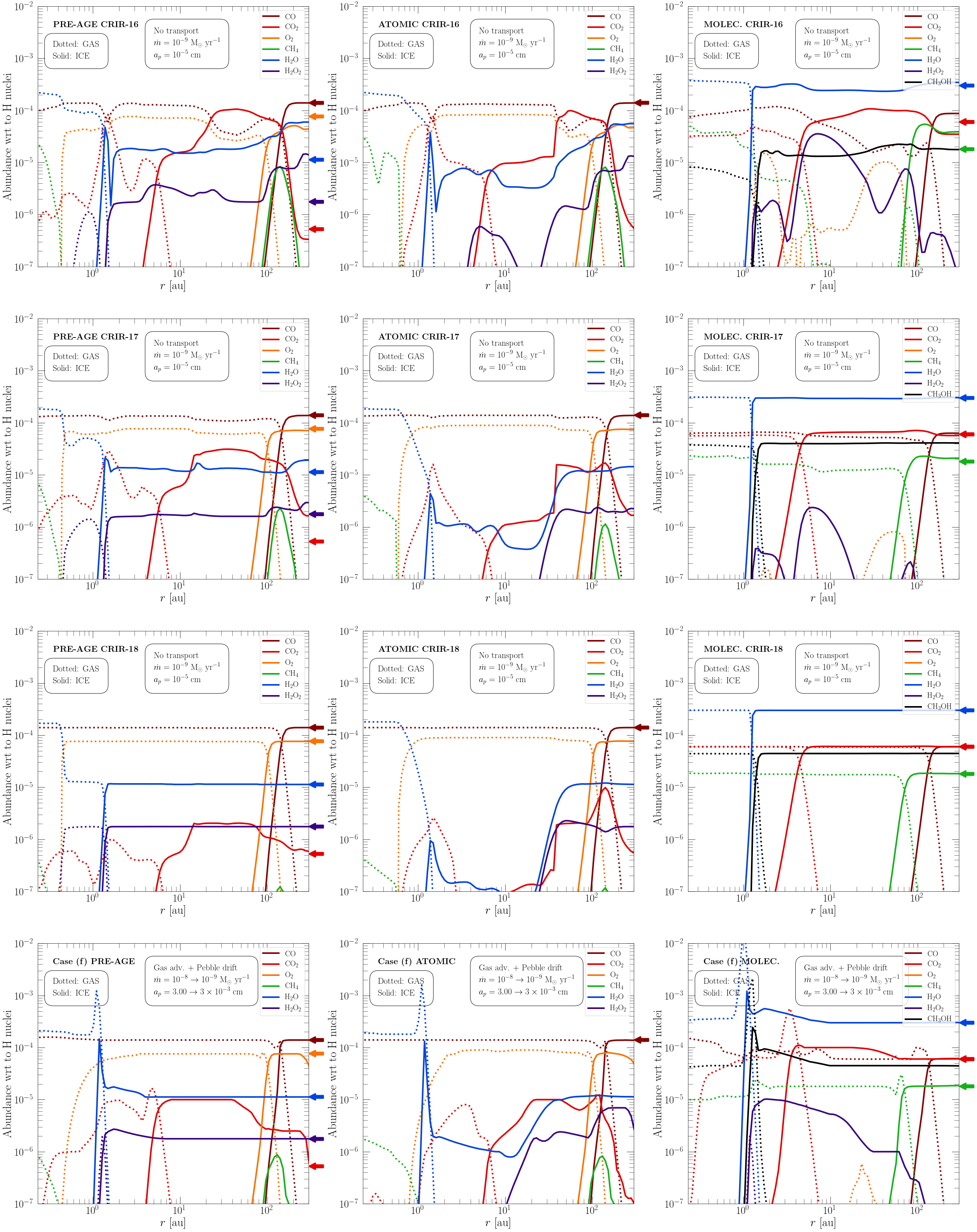}
    \caption{Effect of initial chemical abundances and cosmic ray ionization rate. \textbf{Top three rows:} Chemical evolution after $10^5$ years of the same species as Figure \ref{fig:abu} (plus $\rm CH_3OH$ for the "Molecular" initial condition) for models featuring no transport mechanisms (Case (a)), but different initial conditions and CRIRs (1st row: $10^{-16}\:{\rm s}^{-1}$; 2nd row: $10^{-17}\:{\rm s}^{-1}$; 3rd row: $10^{-18}\:{\rm s}^{-1}$).
    The arrows on right sides of the panels show the initial abundances of the species, which are ice-phase for the "Pre-Age" scenario (1st column), but gas-phase for the "Atomic" (2nd column) and "Molecular" (3rd column) scenarios (see Table \ref{tab:initial}). Note, the PRE-AGE CRIR-17 model is the same as that featured in Figure \ref{fig:abu}a.
    \textbf{Bottom row:} As above, but for Case (f) evolving under different initial/boundary conditions with CRIR of $10^{-17}\:{\rm s}^{-1}$.}
    \label{fig:abu1_NT}
\end{figure*}

\begin{figure*}
    \centering
    \includegraphics[width=0.95\textwidth]{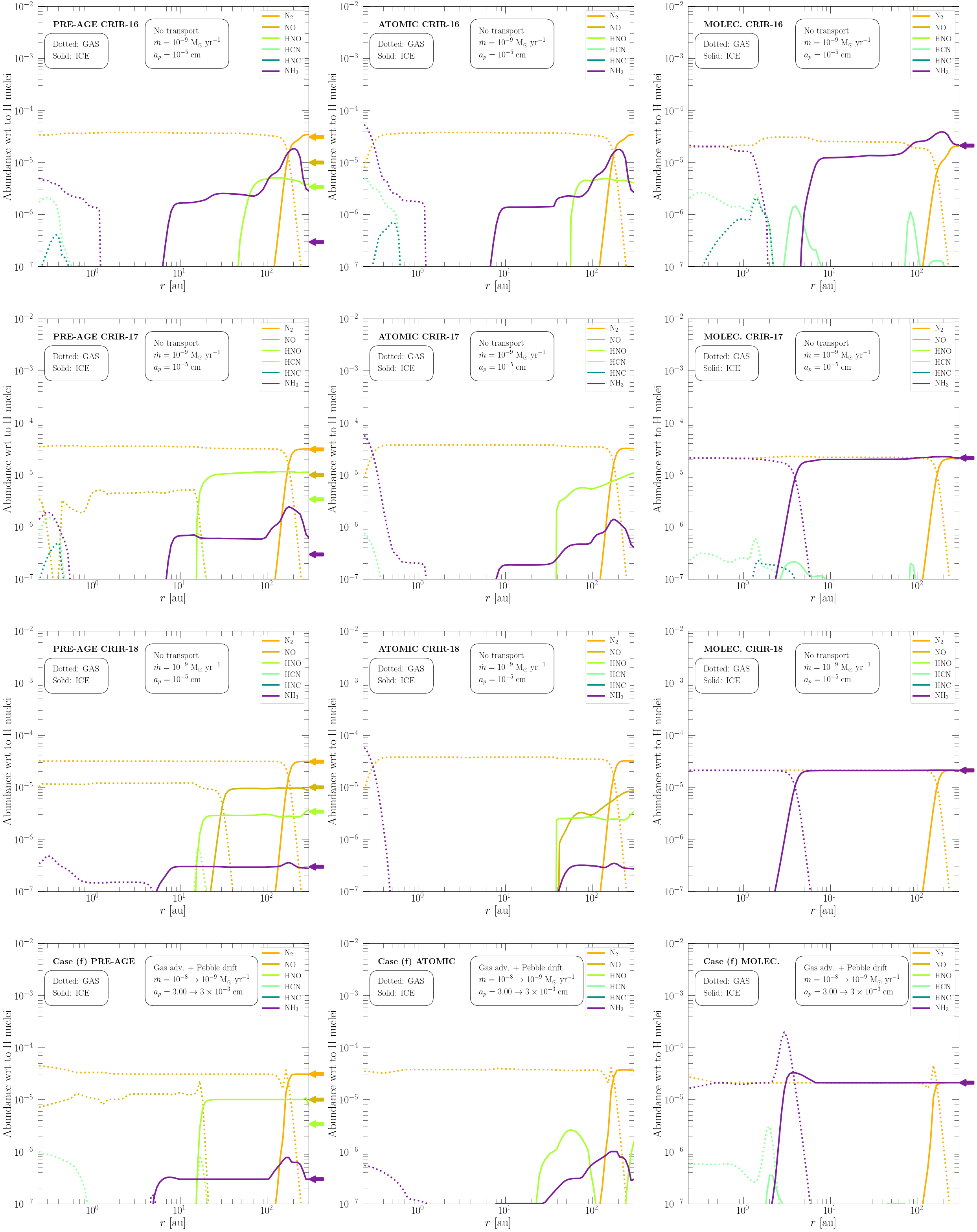}
    \caption{As Fig.~\ref{fig:abu1_NT}, but for nitrogen bearing species.}
    \label{fig:abu2_NT}
\end{figure*}

The first three rows of Figures \ref{fig:abu1_NT} and \ref{fig:abu2_NT} show the chemical evolution of models for Case (a) static disk models, i.e., with physical transport mechanisms turned off, and with $a_p=10^{-5}\:$cm. They are presented as a grid with each column corresponding to different initial/boundary conditions according to those described in Table \ref{tab:initial}; from left to right: Pre-aged, Atomic and Molecular. The Molecular model includes the species $\rm CH_3OH$, which reaches a significant abundance in only this case. Each row corresponds to a different cosmic-ray ionization rate (CRIR); from top to bottom: $\zeta = 10^{-16}$ (CRIR-16), $10^{-17}$ (CRIR-17), and $10^{-18}$ s$^{-1}$ (CRIR-18). The CRIR influences the rate at which chemical species are processed, being particularly necessary for the removal of gas-phase CO, $\rm CH_4$ (important for the molecular case) and NO, to form, with the help of grain-surface chemistry, $\rm H_2O$ and $\rm CO_2$. Therefore, a lower value than our base value of $\zeta = 10^{-17}$ s$^{-1}$ results in smaller changes from the initial composition, while the contrary occurs for a higher value. These results demonstrate the sensitivity of the protoplanetary disk model to CRIR and initial conditions, which influence its composition after $10^{5}$ years.
The bottom rows of both Figures shows the fiducial Case (f) for the corresponding initial/boundary conditions of each column, assuming $\zeta=10^{-17}\:{\rm s}^{-1}$.

{}

{}


\bsp	
\label{lastpage}
\end{document}